\documentclass[a4paper,11pt]{article}
\pdfoutput=1

\usepackage[utf8]{inputenc}
\usepackage[english]{babel}

\usepackage{jinstpub} 
\usepackage{epsfig}
\usepackage{epstopdf}
\usepackage{graphicx}
\usepackage{subfig}
\usepackage{hyperref}

\usepackage{multirow, multicol}
\usepackage{hhline}
\usepackage{makecell}

\title{SiPM-matrix readout of two-phase argon detectors using electroluminescence in the visible and near infrared range}

\collaboration{The DarkSide-20k collaboration}

\newcommand{\Alberta}{Department of Physics, University of Alberta, Edmonton, AB T6G 2R3, Canada}
\newcommand{\APC}{APC, Universit\'e Paris Diderot, CNRS/IN2P3, CEA/Irfu, Obs de Paris, USPC, Paris 75205, France}
\newcommand{\AQLNGS}{INFN Laboratori Nazionali del Gran Sasso, Assergi (AQ) 67100, Italy}
\newcommand{\AQGSSI}{Gran Sasso Science Institute, L'Aquila 67100, Italy}

\newcommand{\Augustana}{Physics Department, Augustana University, Sioux Falls, SD 57197, USA}
\newcommand{\Belgorod}{Radiation Physics Laboratory, Belgorod National Research University, Belgorod 308007, Russia}
\newcommand{\BHSU}{School of Natural Sciences, Black Hills State University, Spearfish, SD 57799, USA}
\newcommand{\BINP}{Budker Institute of Nuclear Physics, Novosibirsk 630090, Russia}
\newcommand{\BNLaddress}{Brookhaven National Laboratory, Upton, NY 11973, USA}
\newcommand{\BOINFN}{INFN Bologna, Bologna 40126, Italy}
\newcommand{\BOUniPHY}{Physics Department, Universit\`a degli Studi di Bologna, Bologna 40126, Italy}
\newcommand{\CAUniCHE}{Department of Mechanical, Chemical, and Materials Engineering, Universit\`a degli Studi, Cagliari 09042, Italy}
\newcommand{\CAUniEEE}{Department of Electrical and Electronic Engineering, Universit\`a degli Studi, Cagliari 09023, Italy}
\newcommand{\CAUniPHY}{Physics Department, Universit\`a degli Studi di Cagliari, Cagliari 09042, Italy}
\newcommand{\CAINFN}{INFN Cagliari, Cagliari 09042, Italy}
\newcommand{\Carleton}{Department of Physics, Carleton University, Ottawa, ON K1S 5B6, Canada}
\newcommand{\Campinas}{Physics Institute, Universidade Estadual de Campinas, Campinas 13083, Brazil}
\newcommand{\CentroFermi}{Museo della fisica e Centro studi e Ricerche Enrico Fermi, Roma 00184, Italy}
\newcommand{\CERNaddress}{CERN, European Organization for Nuclear Research 1211 Geneve 23, Switzerland}
\newcommand{\CIEMAT}{CIEMAT, Centro de Investigaciones Energ\'eticas, Medioambientales y Tecnol\'ogicas, Madrid 28040, Spain}
\newcommand{\Cluj}{National Institute for R\&D of Isotopic and Molecular Technologies, Cluj-Napoca, 400293, Romania}
\newcommand{\CPPM}{Centre de Physique des Particules de Marseille, Aix Marseille Univ, CNRS/IN2P3, CPPM, Marseille, France}

\newcommand{\CTLNS}{INFN Laboratori Nazionali del Sud, Catania 95123, Italy}
\newcommand{\ENUniCEE}{Engineering and Architecture Faculty, Universit\`a di Enna Kore, Enna 94100, Italy}
\newcommand{\ETHZ}{Institute for Particle Physics, ETH Z\"urich, Z\"urich 8093, Switzerland}
\newcommand{\FNALaddress}{Fermi National Accelerator Laboratory, Batavia, IL 60510, USA}
\newcommand{\FortLewis}{Department of Physics and Engineering, Fort Lewis College, Durango, CO 81301, USA}
\newcommand{\GEUni}{Physics Department, Universit\`a degli Studi di Genova, Genova 16146, Italy}
\newcommand{\GEINFN}{INFN Genova, Genova 16146, Italy}

\newcommand{\Hawaii}{Department of Physics and Astronomy, University of Hawai'i, Honolulu, HI 96822, USA}
\newcommand{\Houston}{Department of Physics, University of Houston, Houston, TX 77204, USA}
\newcommand{\IHEPaddress}{Institute of High Energy Physics, Beijing 100049, China}
\newcommand{\IPNO}{Institut de Physique Nucl\`eaire d'Orsay, 91406, Orsay, France}
\newcommand{\INSTM}{Interuniversity Consortium for Science and Technology of Materials, Firenze 50121, Italy}

\newcommand{\JINR}{Joint Institute for Nuclear Research, Dubna 141980, Russia}
\newcommand{\Krakow}{M. Smoluchowski Institute of Physics, Jagiellonian University, 30-348 Krakow, Poland}
\newcommand{\Kurchatov}{National Research Centre Kurchatov Institute, Moscow 123182, Russia}
\newcommand{\Lancaster}{Lancaster University, Lancaster LA1 4YW, UK}
\newcommand{\Laurentian}{Department of Physics and Astronomy, Laurentian University, Sudbury, ON P3E 2C6, Canada}
\newcommand{\LNFINFN}{INFN Laboratori Nazionali di Frascati, Frascati 00044, Italy}

\newcommand{\Lodz}{Institute of Applied Radiation Chemistry, Lodz University of Technology, 93-590 Lodz, Poland}
\newcommand{\LPNHE}{LPNHE, CNRS/IN2P3, Sorbonne Universit\'e, Universit\'e Paris Diderot, Paris 75252, France}

\newcommand{\Manchester}{The University of Manchester, Manchester M13 9PL, UK}
\newcommand{\MEPhI}{National Research Nuclear University MEPhI, Moscow 115409, Russia}

\newcommand{\MIINFN}{INFN Milano, Milano 20133, Italy}
\newcommand{\MIPoliICA}{Civil and Environmental Engineering Department, Politecnico di Milano, Milano 20133, Italy}
\newcommand{\MIPoliCHE}{Chemistry, Materials and Chemical Engineering Department ``G.~Natta", Politecnico di Milano, Milano 20133, Italy}
\newcommand{\MIPoliEIB}{Electronics, Information, and Bioengineering Department, Politecnico di Milano, Milano 20133, Italy}
\newcommand{\MIPoliENE}{Energy Department, Politecnico di Milano, Milano 20133, Italy}
\newcommand{\MIUni}{Physics Department, Universit\`a degli Studi di Milano, Milano 20133, Italy}
\newcommand{\MSU}{Skobeltsyn Institute of Nuclear Physics, Lomonosov Moscow State University, Moscow 119234, Russia}
\newcommand{\NAINFN}{INFN Napoli, Napoli 80126, Italy}
\newcommand{\NAUniPHY}{Physics Department, Universit\`a degli Studi ``Federico II'' di Napoli, Napoli 80126, Italy}
\newcommand{\NAUniCHE}{Chemical, Materials, and Industrial Production Engineering Department, Universit\`a degli Studi ``Federico II'' di Napoli, Napoli 80126, Italy}
\newcommand{\NAUniPHARM}{Pharmacy Department, Universit\`a degli Studi ``Federico II'' di Napoli, Napoli 80131, Italy}
\newcommand{\NAUniEEIT}{Department of Electrical Engineering and Information Technology, Universit\`a degli Studi ``Federico II'' di Napoli, Napoli 80125, Italy}
\newcommand{\NSU}{Novosibirsk State University, Novosibirsk 630090, Russia}

\newcommand{\Petersburg}{Saint Petersburg Nuclear Physics Institute, Gatchina 188350, Russia}
\newcommand{\PGUniCBB}{Chemistry, Biology and Biotechnology Department, Universit\`a degli Studi di Perugia, Perugia 06123, Italy}
\newcommand{\PGINFN}{INFN Perugia, Perugia 06123, Italy}
\newcommand{\PIINFN}{INFN Pisa, Pisa 56127, Italy}
\newcommand{\PIUniPHY}{Physics Department, Universit\`a degli Studi di Pisa, Pisa 56127, Italy}
\newcommand{\PNNLaddress}{Pacific Northwest National Laboratory, Richland, WA 99352, USA}
\newcommand{\Princeton}{Physics Department, Princeton University, Princeton, NJ 08544, USA}
\newcommand{\Queens}{Department of Physics, Engineering Physics and Astronomy, Queen's University, Kingston, ON K7L 3N6, Canada}
\newcommand{\RHUL}{Department of Physics, Royal Holloway University of London, Egham TW20 0EX, UK}
\newcommand{\RMTreINFN}{INFN Roma Tre, Roma 00146, Italy}
\newcommand{\RMTreUni}{Mathematics and Physics Department, Universit\`a degli Studi Roma Tre, Roma 00146, Italy}
\newcommand{\RMUnoINFN}{INFN Sezione di Roma, Roma 00185, Italy}
\newcommand{\RMUnoUni}{Physics Department, Sapienza Universit\`a di Roma, Roma 00185, Italy}
\newcommand{\SAINFN}{INFN Salerno, Salerno 84084, Italy}
\newcommand{\SAUniPHY}{Physics Department, Universit\'a degli Studi di Salerno, Salerno 84084, Italy}
\newcommand{\SNOLABaddress}{SNOLAB, Lively, ON P3Y 1N2, Canada}
\newcommand{\SSUniCHP}{Chemistry and Pharmacy Department, Universit\`a degli Studi di Sassari, Sassari 07100, Italy}
\newcommand{\Sussex}{Physics and Astronomy, University of Sussex, Brighton BN1 9QH, UK}

\newcommand{\TOINFN}{INFN Torino, Torino 10125, Italy}
\newcommand{\TOPoli}{Department of Electronics and Communications, Politecnico di Torino, Torino 10129, Italy}

\newcommand{\TRIUMFaddress}{TRIUMF, 4004 Wesbrook Mall, Vancouver, BC V6T 2A3, Canada}
\newcommand{\TUM}{Physik Department, Technische Universit\"at M\"unchen, Munich 80333, Germany}
\newcommand{\UB}{Universitat de Barcelona, Barcelona E-08028, Catalonia, Spain} 
\newcommand{\UCDavis}{Department of Physics, University of California, Davis, CA 95616, USA}
\newcommand{\UCLA}{Physics and Astronomy Department, University of California, Los Angeles, CA 90095, USA}
\newcommand{\UMass}{Amherst Center for Fundamental Interactions and Physics Department, University of Massachusetts, Amherst, MA 01003, USA}
\newcommand{\UNAM}{Instituto de F\'isica, Universidad Nacional Aut\'onoma de M\'exico (UNAM), M\'exico 01000, Mexico}
\newcommand{\UOC}{Department of Chemistry, University of Crete, P.O. Box 2208, 71003 Heraklion, Crete, Greece}
\newcommand{\USP}{Instituto de F\'isica, Universidade de S\~ao Paulo, S\~ao Paulo 05508-090, Brazil}
\newcommand{\VTech}{Virginia Tech, Blacksburg, VA 24061, USA}
\newcommand{\Zaragoza}{Centro de Astropart\'iculas y F\'isica de Altas Energ\'ias, Universidad de Zaragoza, Zaragoza 50009, Spain}
\newcommand{\ARAID}{ARAID, Fundaci\'on Agencia Aragonesa para la Investigaci\'on y el Desarrollo, Gobierno de Arag\'on, Zaragoza 50018, Spain}

\author[1]{C.~E.~Aalseth,}
\author[2]{S.~Abdelhakim,}
\author[3]{P.~Agnes,}
\author[4]{R.~Ajaj,}
\author[5]{I.~F.~M.~Albuquerque,}
\author[1]{T.~Alexander,}
\author[6,7]{A.~Alici,}
\author[8]{A.~K.~Alton,}
\author[9]{P.~Amaudruz,}
\author[10]{F.~Ameli,}
% ...
\affiliation[1]{\PNNLaddress}
\affiliation[2]{\IPNO}
\affiliation[3]{\Houston}
\affiliation[4]{\Carleton}
\affiliation[5]{\USP}
\affiliation[6]{\BOUniPHY}
\affiliation[7]{\BOINFN}
\affiliation[8]{\Augustana}
\affiliation[9]{\TRIUMFaddress}
\affiliation[10]{\RMUnoINFN}
% ...

\emailAdd{ds-ed@lngs.infn.it}

\abstract{Proportional electroluminescence (EL) in noble gases is used in two-phase detectors for dark matter searches to record (in the gas phase) the ionization signal induced by particle scattering in the liquid phase. The ``standard'' EL  mechanism is considered to be due to noble gas excimer emission in the vacuum ultraviolet (VUV). In addition, there are two alternative mechanisms, producing light in the visible and near infrared (NIR) ranges. The first is due to bremsstrahlung of electrons scattered on neutral atoms (``neutral bremsstrahlung'', NBrS). The second, responsible for electron avalanche scintillation in the NIR at higher electric fields, is due to transitions between excited atomic states. In this work, we have for the first time demonstrated two alternative techniques of the optical readout of two-phase argon detectors, in the visible and NIR range, using a silicon photomultiplier matrix and electroluminescence due to either neutral bremsstrahlung or avalanche scintillation. The amplitude yield and position resolution were measured for these readout techniques, which allowed to assess the detection threshold for electron and nuclear recoils in two-phase argon detectors for dark matter searches. 
	To the best of our knowledge, this is the first practical application of the NBrS effect in detection science.}

\begin{document}

\maketitle
\flushbottom
\newcommand*{\doi}[1]{\href{http://dx.doi.org/#1}{doi: #1}}

\section{Introduction}
In two-phase noble liquid detectors for dark matter searches and low-energy neutrino experiments, the scattered particle produces two types of signals~\cite{Chepel2013}: that of primary scintillation, recorded promptly (``S1'') and that of primary ionization, which is recorded with a delay (``S2'').
The S1 and S2 signals are typically recorded by photomultiplier tubes (PMTs)~\cite{Agnes2015} or silicon photomultiplier (SiPM) matrices~\cite{Aalseth2018}, adapted for operation at cryogenic temperatures.

The detection of S1 serves to find the z coordinate of the interaction vertex (by the S1-S2 time difference). In addition, it is very useful for effective selection between nuclear and electron recoils, using either the S1 pulse-shape or the S1/S2 ratio for discrimination. However, the S1 detection is not a must: in particular, at lower recoil energies where the S1 signal is too weak, experiments can resort to an S2-only operation mode~\cite{Agnes2018,Aprile2019_XENON1T}. Accordingly, in this paper we confine ourselves to the issue of S2-only signal detection, leaving issues related to S1 to future studies.

S2 detection in dual-phase detectors relies on proportional electroluminescence (EL) in the noble gas~\cite{Oliveira2011,Buzulutskov2020}.
In argon, the ordinary (``standard'') mechanism of proportional electroluminescence is considered to be due to vacuum ultraviolet (VUV) emission (around 128~nm) of noble gas excimers Ar$^{*}_{2}$($^{1,3}\Sigma^{+}_{u}$) produced in three-body atomic collisions of excited atoms Ar$^*$(3p$^5$4s$^1$), which in turn are produced by drifting electrons in electron-atom collisions: see review~\cite{Buzulutskov2020}.

While present two-phase argon detectors rely on PMTs, the baseline design option for the future dark matter detector DarkSide-20k~\cite{Aalseth2018} employs SiPM matrices to record ordinary electroluminescence in the VUV. The sensitivity of PMTs and SiPMs is limited to the visible or NUV range~\cite{Aalseth2017,Acerbi2017}.
The development of SiPMs with VUV sensitivity, namely at 128~nm, is still at the R\&D stage~\cite{Igarashi2016}, since such SiPMs are windowless~\cite{Ghassemi2015} which limits their use in cryogenic conditions and over large sensitive areas.
It is thus necessary to convert the VUV into visible light using a wavelenght shifter (WLS), namely tetraphenyl-butadiene (TPB). 
An issue with TPB is that it may not be stable over long time scales, in particular due to its dissolving in liquid Ar~\cite{Asaadi2019} and peeling off from the substrate under cryogenic conditions~\cite{Burak2020}. Another known issue is related to difficulties in achieving uniform levels of WLS deposits over large detector areas.

Alternative readout techniques for two-phase argon detector, proposed elsewhere~\cite{Buzulutskov2011,Buzulutskov2018} and based on ``non-standard'' electroluminescence in the visible and near infrared (NIR) range, could allow detector operation without WLS.
In this work, we demonstrate the successful S2-only performance of a two-phase argon detector with SiPM-matrix optical readout, in the visible and NIR range, using  two such alternative readout techniques.

This study was performed using the experimental setup of the Novosibirsk group of the DarkSide collaboration.

\section{Alternative concepts of SiPM-matrix readout of two-phase argon detectors}\label{Two_concepts}

\begin{figure}[ht!]
	\center{\includegraphics[width=0.6\columnwidth]{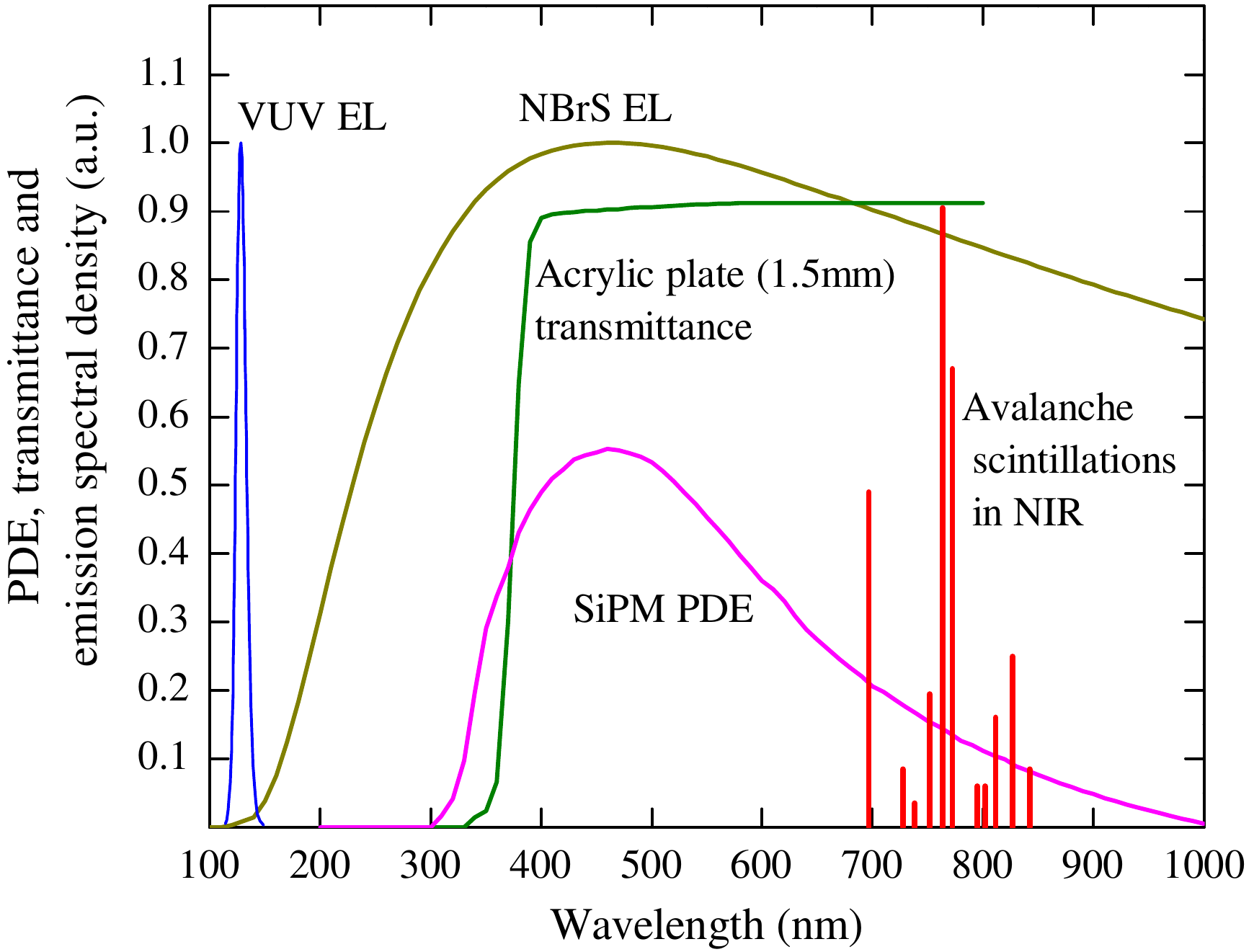}}
	\caption{Photon emission spectra in gaseous Ar due to ordinary scintillations in the VUV measured in \cite{Morozov2008}, NBrS electroluminescence at 8.3~Td theoretically calculated in \cite{Buzulutskov2018} and avalanche scintillations in the NIR measured in \cite{Lindblom1988,Fraga2000}. Also shown are the photon detection efficiency (PDE) of the SiPMs (MPPC 13360-6050PE~\cite{hamamatsu}) at an overvoltage of 5.6~V obtained from~\cite{Otte2017} and the transmittance of the acrylic plate (1.5~mm thick) in front of the SiPM matrix, used in this study}
	\label{image:fig_optical_spectra}
\end{figure}

In argon, ordinary electroluminescence (in the VUV, around 128 nm, see Figure~\ref{image:fig_optical_spectra}) goes via Ar$^*$(3p$^5$4s$^1$) atomic excited states~\cite{Buzulutskov2020} and thus has a threshold for the reduced electric field of about 4~Td~\cite{Buzulutskov2018}, which is defined by the energy threshold for Ar atom excitation. 
The reduced electric field is defined as $\mathcal{E}/N$ expressed in Td units (1~Td~=~$10^{-17}$~V~cm$^2$) corresponding to 0.87~kV/cm in gaseous argon at a pressure of 1.00~atm and a temperature of 87.3~K, where $\mathcal{E}$ is the electric field and $N$ is the atomic density.

In addition to the ordinary EL mechanism, a concurrent EL mechanism, based on bremsstrahlung of drifting electrons scattered on neutral atoms (so-called ``neutral bremsstrahlung'', NBrS), has been recently revealed~\cite{Buzulutskov2018,Bondar2019}. It was shown that the NBrS effect can explain two remarkable properties of proportional electroluminescence: photon emission below the Ar excitation threshold and the substantial contribution of a non-VUV spectral component. NBrS electroluminescence has a continuous emission spectrum, extending from the UV to the visible and NIR range: see~Figure~\ref{image:fig_optical_spectra}. 

At higher electric fields (above 8~Td), another ``non-standard'' EL mechanism comes into force, namely that of electroluminescence in the NIR due to transitions between excited atomic states~\cite{Lindblom1988,Fraga2000, Buzulutskov2011, StudyInfraredScintillations2012P2,Oliveira2013,Buzulutskov2020}: Ar$^{*}$(3p$^5$4p$^{1}$)$\longrightarrow$Ar$^{*}$(3p$^5$4s$^{1}$). It has a line emission spectrum in the range of 700 to 850~nm (Figure~\ref{image:fig_optical_spectra}). Similarly to the ordinary mechanism, the excited Ar$^{*}$(3p$^5$4p$^{1}$) atoms are produced by drifting electrons in electron-atom collisions.
This mechanism is particularly noticeable at even higher fields, above 30 Td, where avalanche multiplication takes place, accompanied by secondary scintillation(``avalanche scintillation'')~\cite{Fraga2000,Bondar2010}.

\begin{figure}[ht!]
	\center{\includegraphics[width=0.6\columnwidth]{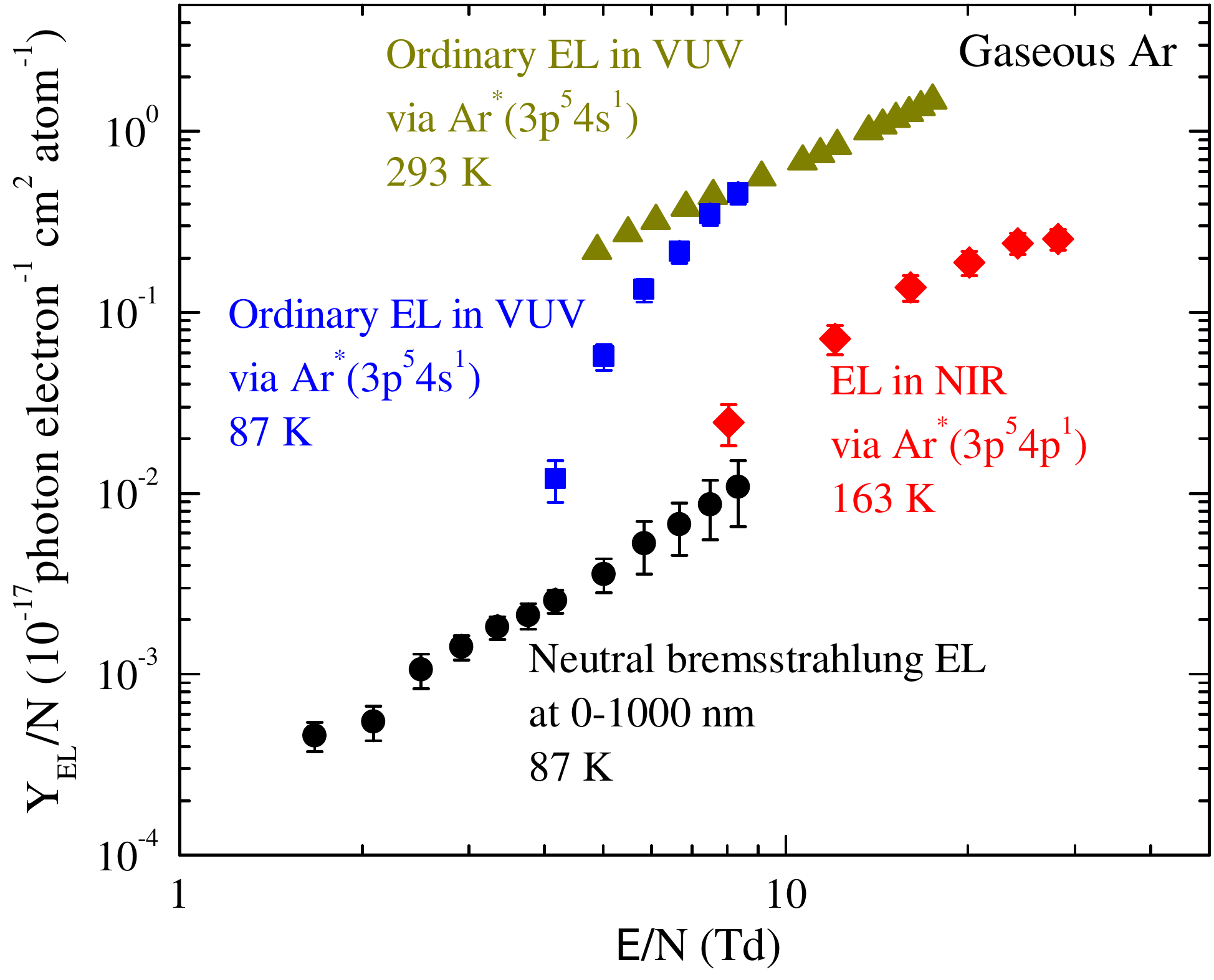}}
	\caption{Summary of experimental data on the reduced EL yield in argon for all known electroluminescence (EL) mechanisms: for NBrS EL at wavelengths of 0-1000 nm, measured in~\cite{Bondar2019} at 87~K; for ordinary EL in the VUV, going via Ar$^*$(3p$^5$4s$^1$), measured in~\cite{Bondar2019} at 87~K and in~\cite{Monteiro2008} at 293~K; for EL in the NIR going via Ar$^*$(3p$^5$4p$^1$), measured in~\cite{Buzulutskov2011} at 163~K}
	\label{fig_reduced_EL_for_all_mech}
\end{figure}

Figure~\ref{fig_reduced_EL_for_all_mech} presents all known experimental data on reduced EL yield in argon for all known EL mechanisms:
for NBrS electroluminescence at wavelengths below 1000 nm, for ordinary electroluminescence in the VUV and for electroluminescence in the NIR. In addition, Figure~\ref{image:fig_optical_spectra} shows their photon emission spectra, along with the spectral response of the SiPMs used in the present study.

\begin{figure}[ht!]
	\center{\includegraphics[width=0.4\columnwidth]{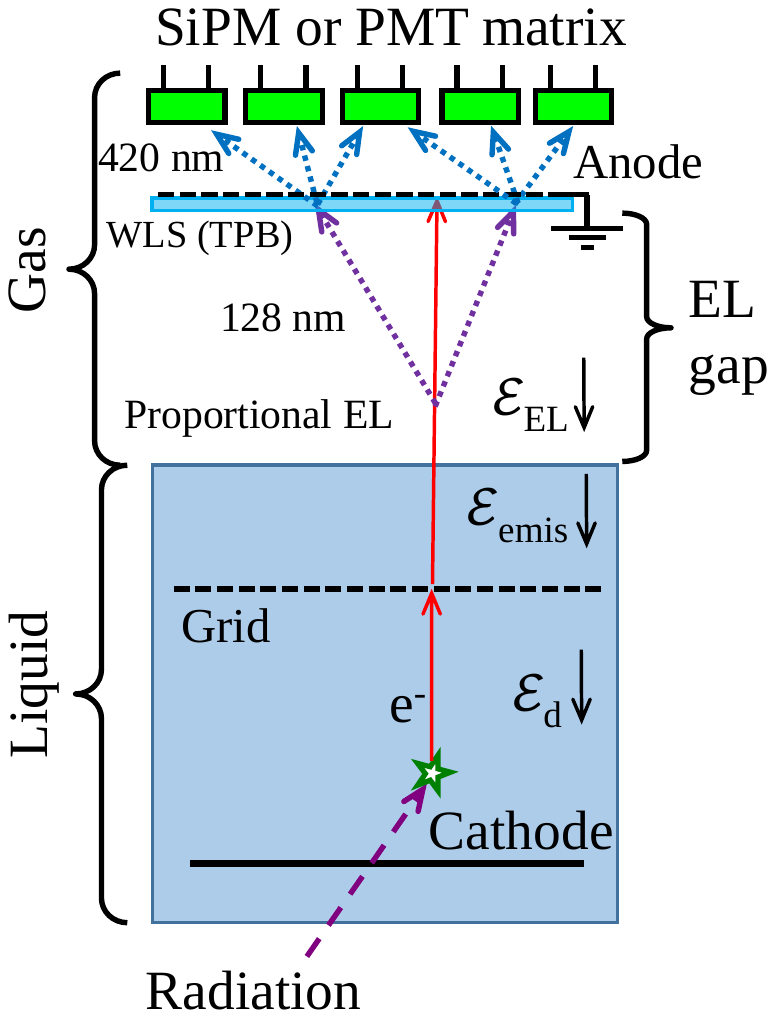}}
	\caption{``Standard'' concept of SiPM-matrix readout of two-phase argon detectors with an EL gap}
	\label{fig_conceptual_scheme_ordinary_EL}
\end{figure}

The ``standard'' concept of SiPM matrix readout of two-phase argon detectors is depicted in Figure~\ref{fig_conceptual_scheme_ordinary_EL}. In this concept the SiPM matrix is coupled to the EL gap via a wavelength shifter (WLS).

\begin{figure}[ht!]
	\centering
	\subfloat{\includegraphics[width=0.4\columnwidth]{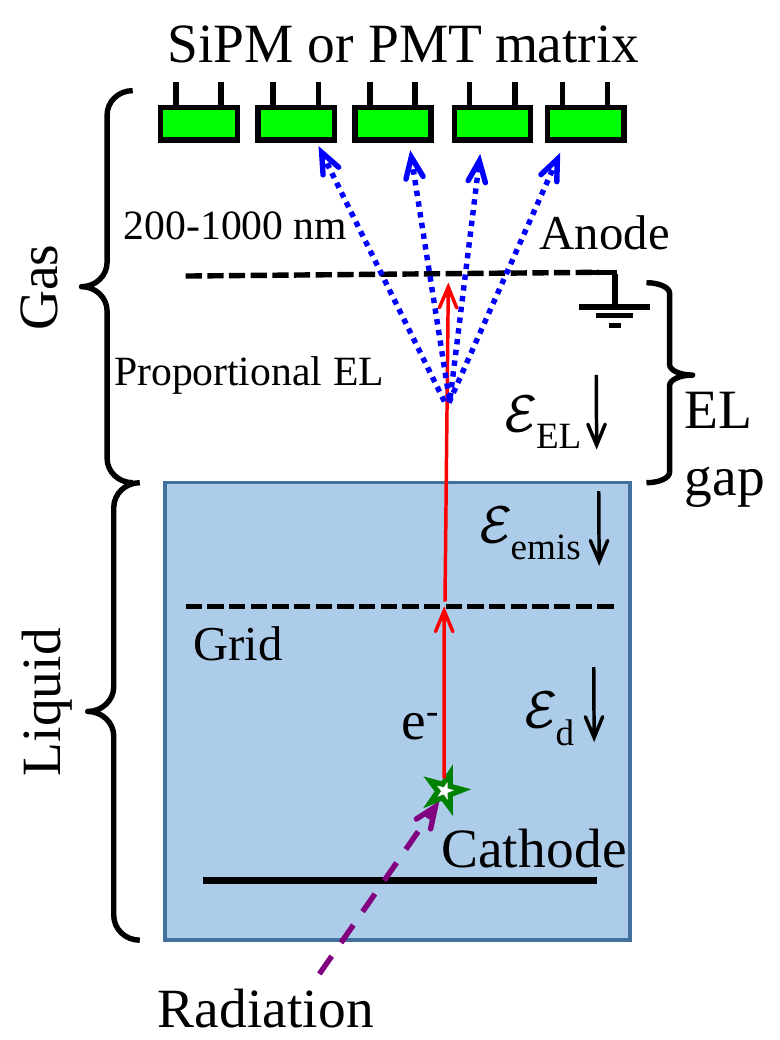}\label{fig_conceptual_scheme_NBrS_EL}}
	\subfloat{\includegraphics[width=0.4\columnwidth]{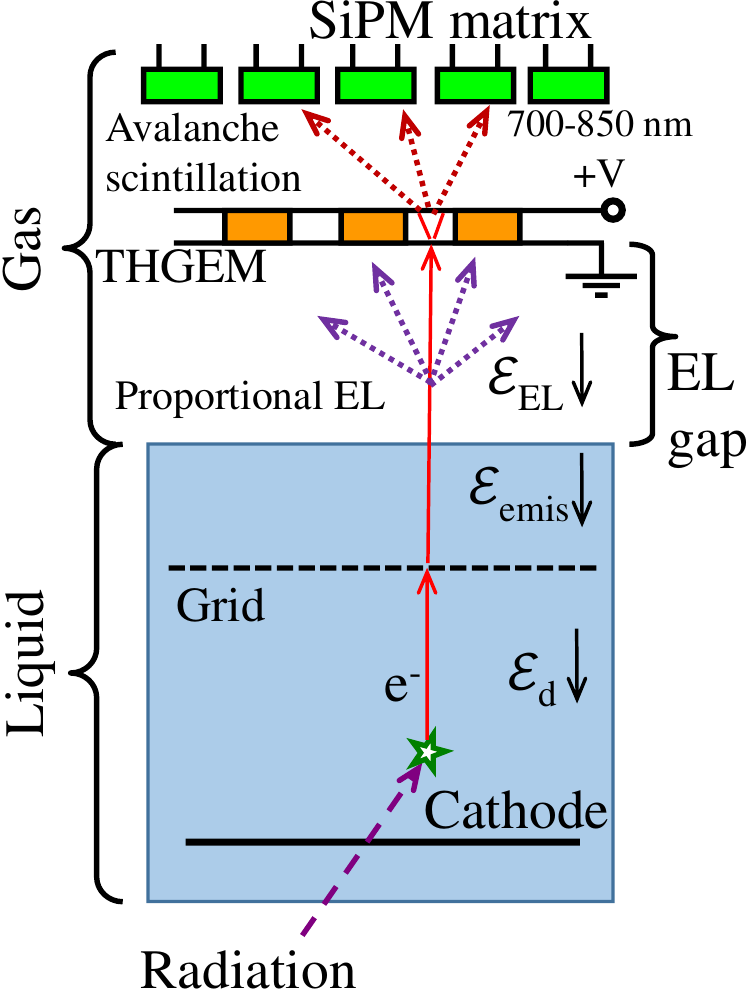}\label{fig_conceptual_scheme_avalanche}} 
	\caption[] { 
		Two alternative concepts of SiPM-matrix readout of two-phase argon detectors with EL gap proposed elsewhere \cite{Buzulutskov2011,Buzulutskov2018} and experimentally studied in the present work: that of SiPM matrix directly coupled to EL gap  (``direct SiPM-matrix readout'') (left) and that of combined THGEM/SiPM-matrix multiplier coupled to EL gap (``THGEM/SiPM-matrix readout'') (right)}
	\label{fig_conceptual_scheme}
\end{figure}

Figure~\ref{fig_conceptual_scheme} illustrates two alternative readout concepts proposed elsewhere ~\cite{Buzulutskov2011,StudyInfraredScintillations2012P2,Buzulutskov2018} and realized in the present study. These are based on NBrS electroluminescence and avalanche scintillations in the NIR, respectively. 

In the first alternative concept~\cite{Buzulutskov2018}, the EL gap is read out directly in the visible and NIR range, using a SiPM matrix directly coupled to the EL gap. In the second alternative concept~\cite{Buzulutskov2011,StudyInfraredScintillations2012P2}, the EL gap is read out indirectly, using a combined THGEM/SiPM-matrix multiplier coupled to the EL gap, the THGEM being operated in electron avalanche mode. The advantage of these concepts is operating without a WLS. As noted above, this may lead to more stable operation of two-phase argon detectors, avoiding the problems of WLS degradation and its dissolving in liquid Ar~\cite{Asaadi2019}, as well as that of WLS peeling off from the substrate. 

In the first alternative concept, hereinafter referred to as ``direct SiPM-matrix readout'', the detection threshold for S2 signal might increase compared to that of the ``standard'' concept at higher electric fields (exceeding 5~Td), since here the light yield of NBrS electroluminescence is lower compared to that of ordinary electroluminescence: see Figure~\ref{fig_reduced_EL_for_all_mech}. On the other hand, for lower reduced electric fields, between 4 and 5~Td, 
the response of PMTs and SiPMs to NBrS electroluminescence might be comparable with that of ordinary electroluminescence recorded using a WLS~\cite{Buzulutskov2018}. 
This is because the NBrS EL is recorded mostly directly thanks to its spectrum part in the visible range, and thus practically without losses,  while ordinary EL is recorded indirectly, with significant reduction of the photon flux  after re-emission by the WLS, the reduction factor reaching 15-20~\cite{Buzulutskov2018} (in the absence of optical contact between the WLS and SiPM).
Note that the EL field employed in the DarkSide-50 search for low-mass WIMPs in an S2-only mode was 4.2 kV/cm, corresponding to 4.8 Td~\cite{Agnes2018}.

In the second alternative concept, hereinafter referred to as ``THGEM/SiPM-matrix readout'', an additional charge amplification of the S2 signal is provided by applying a voltage across the THGEM, resulting in electron avalanching in THGEM holes. Accordingly, the SiPM matrix records avalanche scintillations in the NIR from the THGEM holes, rather than electroluminescence from the EL gap. In this case, the detection threshold for the S2 signal can be significantly decreased, compared to direct SiPM-matrix readout.

It should be remarked that the concept of THGEM/SiPM-matrix readout overlaps with the earlier idea of Cryogenic Avalanche Detectors (CRADs), developed elsewhere~\cite{Buzulutskov2012,Buzulutskov2020,THGEM_GAPD_optical_readout_2013}. In CRADs, the charge multiplication or avalanche scintillation signal from the THGEM (or GEM), placed in the gas phase of the two-phase detector, is recorded. The difference is that in the CRAD concept, the gas gap underneath the THGEM is not supposed to operate in EL mode (i.e. it operates at relatively low electric fields). In contrast, in THGEM/SiPM-matrix readout concept, the EL gap is needed to record proportional electroluminescence in addition to that of avalanche scintillation, using either the bottom or side SiPM matrices, to provide the superior amplitude resolution when recording single drifting electrons. This is needed because in these conditions the amplitude resolution of the THGEM is not sufficient: it is significantly deteriorated due to intrinsic fluctuations of the electron avalanche~\cite{Bondar2008}.

In the following sections, we first describe the development of SiPM matrices for operation in two-phase argon detectors, and then report the implementation of these alternative readout concepts in our experimental setup.

\section{R$\&$D of SiPM matrices operated in two-phase argon detectors}\label{Choice_of_SiPMs}
In the course of this study, three SiPM matrices were progressively developed for operation in two-phase argon detectors, with a channel pitch of 1~cm and matrix size of 5$\times$5 of active channels. Three different types of SiPMs were used in the matrices, respectively: see Table~\ref{table:SiPM_characteristics}. 

\begin{table}
	\centering
	\caption{Characteristic properties of three SiPMs types used in SiPM matrices}
	\label{table:SiPM_characteristics}
	\begin{tabular}{@{\extracolsep{\fill}}|cccc|@{}}
		\hline
		\makecell[c]{SiPM type \\ (producer)} & \makecell{Active \\ area (mm$^2$)} & \makecell{Number of \\ pixels } & \makecell{Active area \\ fill factor (\%)} \\
		\hline
		\rule{0pt}{7ex}
		\makecell[c]{MRS APD \\ 149-35 \\ (CPTA)} & 2.1$\times$2.1 & 1764 & 62 \\  
		\makecell[c]{MPPC \\ S10931-100P \\ (Hamamatsu)}  & 3$\times$3 & 900 & 78.5 \rule{0pt}{7ex} \\
		\makecell[c]{MPPC \\ S13360-6050PE \\ (Hamamatsu)}  & 6$\times$6 & 14400 & 74 \rule{0pt}{7ex} \\
		\hline
	\end{tabular}
\end{table}

\begin{figure}
	\centering
	\subfloat[]{\includegraphics[width=0.4\columnwidth]{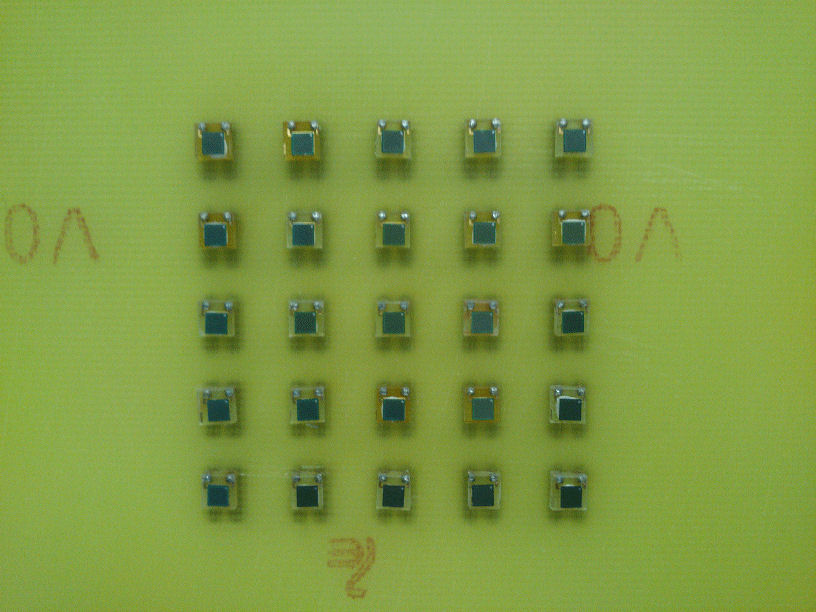}\label{image:fig_5x5el_CPTA_2x2mm2}} \hspace{5mm}
	\subfloat[]{\includegraphics[width=0.4\columnwidth]{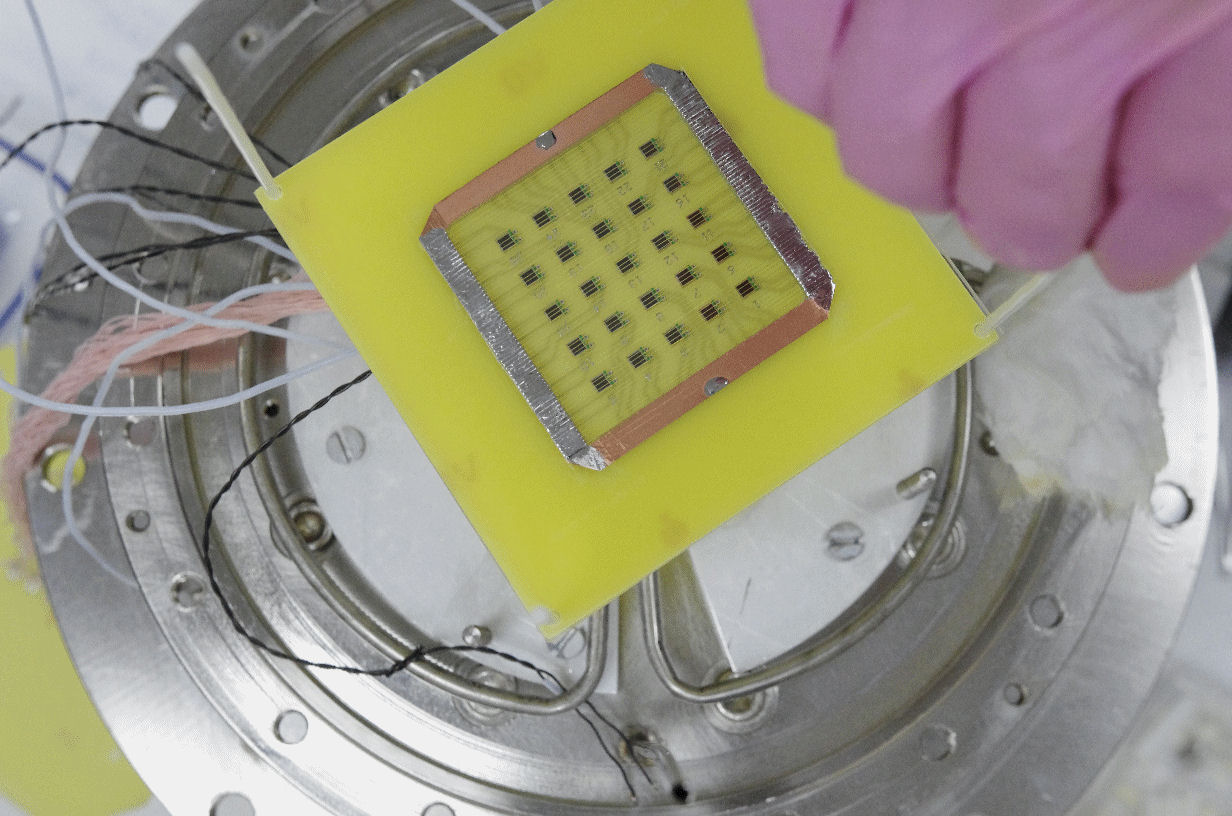}\label{image:fig_5x5el_Ham_3x3mm2}} \\
	\subfloat[]{\includegraphics[width=0.4\columnwidth]{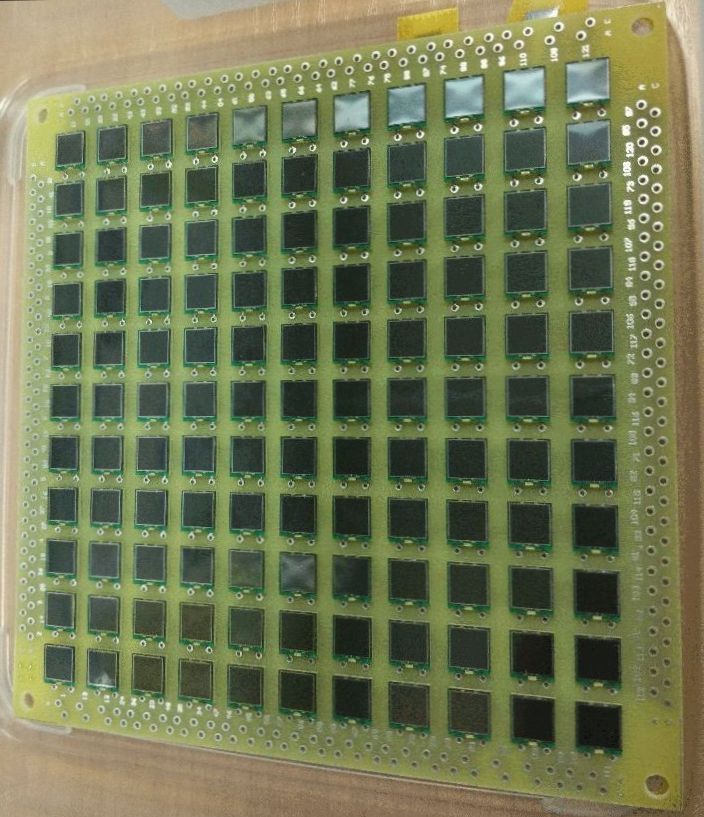}\label{image:fig_SiPM_11x11_matrix_photo}}
	\caption[] {\label{image:fig_all_SiPM_matrices_photo}  Photographs of SiPM matrices progressively developed in this work. \subref{image:fig_5x5el_CPTA_2x2mm2} 5$\times$5 SiPM matrix made from MRS APD 149-35 (CPTA) with an active area of 2.1$\times$2.1~mm$^2$.
		\subref{image:fig_5x5el_Ham_3x3mm2} 5$\times$5 SiPM matrix made from MPPC S10931-100P (Hamamatsu) with an active area of 3$\times$3~mm$^2$.
		\subref{image:fig_SiPM_11x11_matrix_photo} 11$\times$11 SiPM matrix made from MPPC S13360-6050PE (Hamamatsu) with an active area of 6$\times$6~mm$^2$. The SiPM channel pitch is 1~cm in all cases
	} 	
\end{figure}

\begin{figure}[ht!]
	\center{\includegraphics[width=0.65\columnwidth]{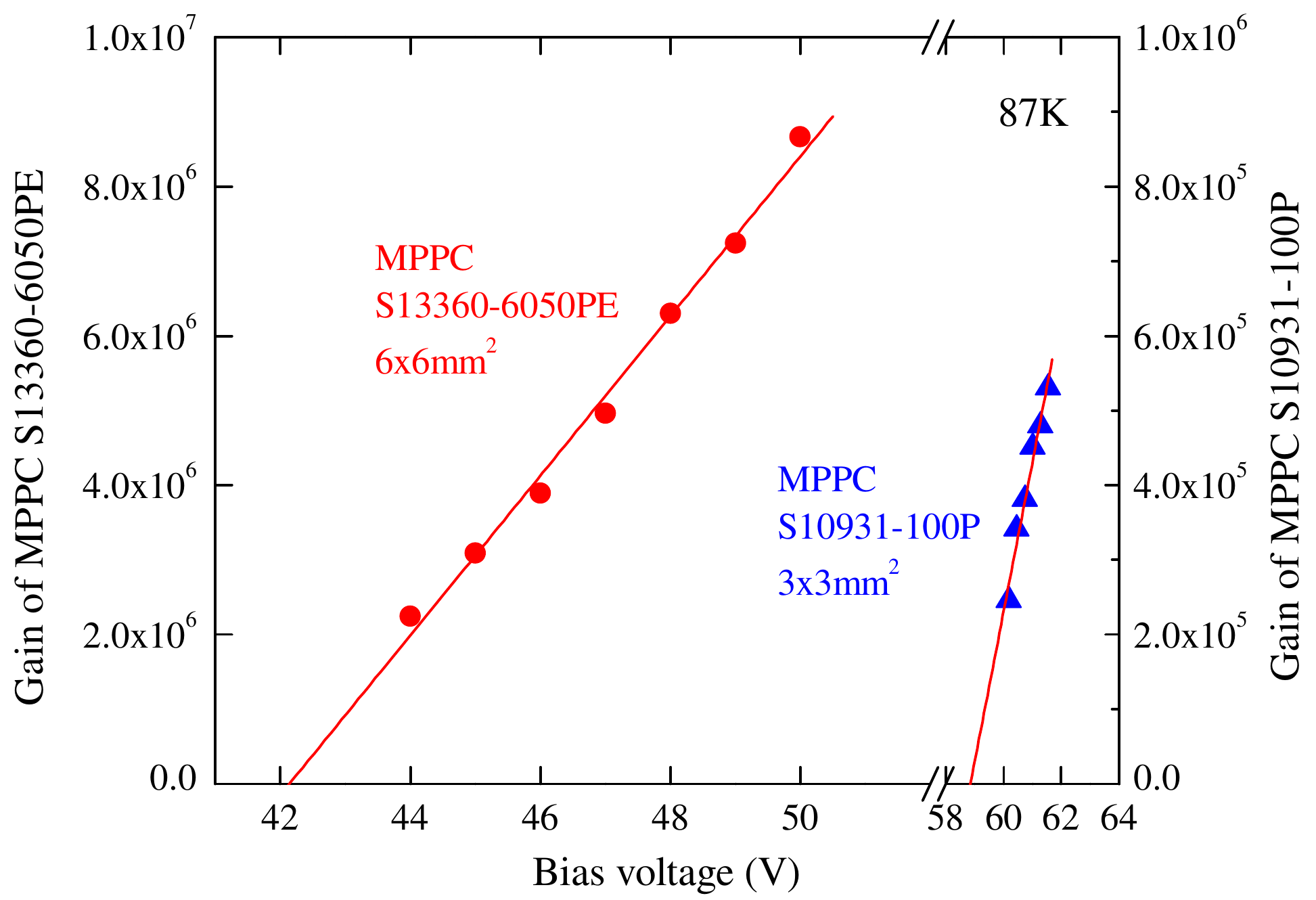}}
	\caption{Gain-voltage characteristics of different SiPM types at 87~K}
	\label{image:fig_gain_CPTA_149-35__MPPC_S10931-100P__S13360-6050PE}
\end{figure}

\begin{figure}[ht!]
	\center{\includegraphics[width=0.55\columnwidth]{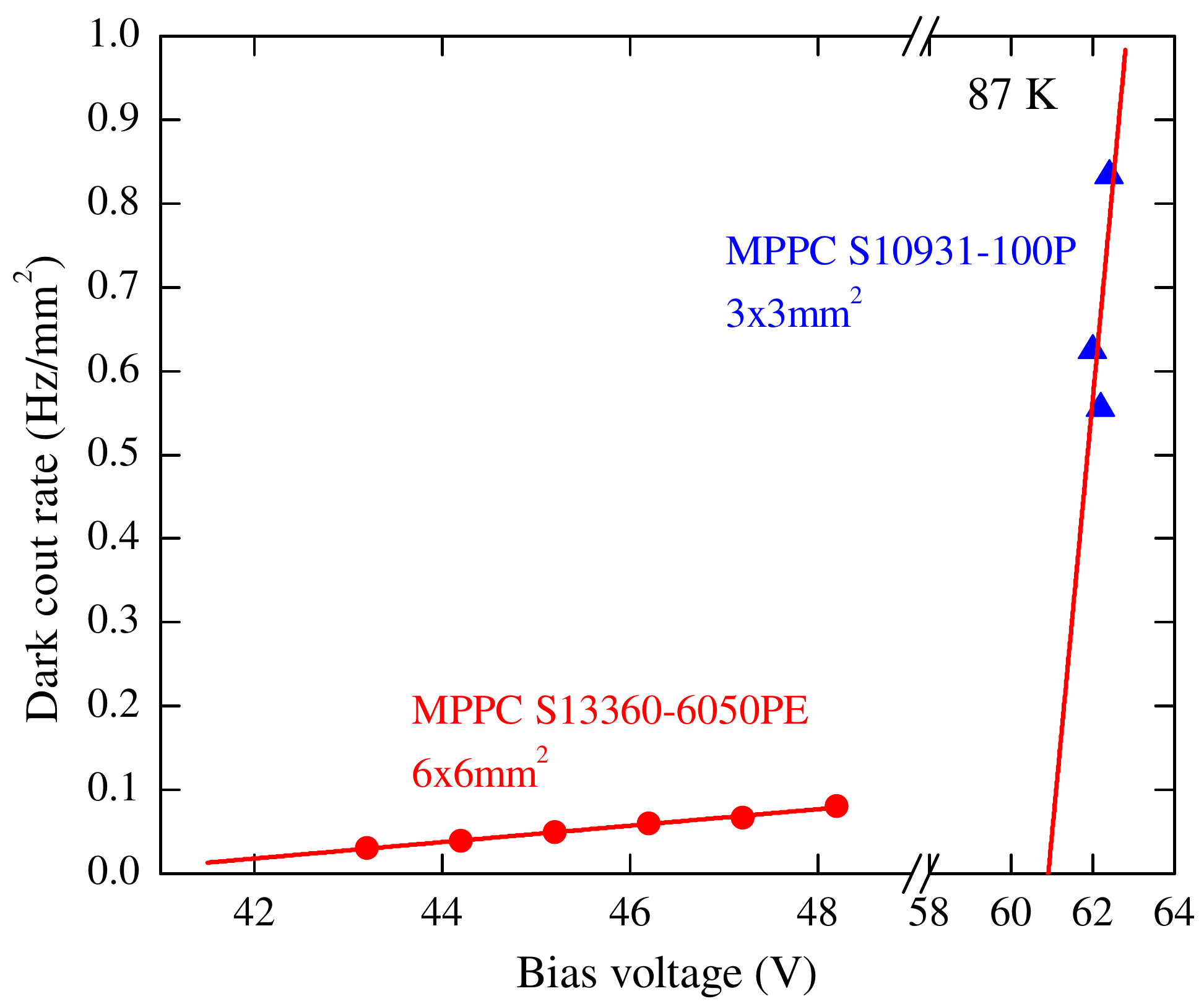}}
	\caption{Dark count rates of different SiPM types as a function of the bias voltage at 87~K}
	\label{image:fig_noise_CPTA_149-35__MPPC_S10931-100P__S13360-6050PE}
\end{figure}

The first SiPM type was MRS APD 149-35 (CPTA)~\cite{CPTA_company}: see Figure~\ref{image:fig_5x5el_CPTA_2x2mm2}. At 87~K, it showed an acceptable dark count rate (about 6~Hz/mm$^2$) with a gain of about 10$^6$~\cite{Bondar2011}. However, during the first cryogenic run, half of the 25 channels failed, making impossible further use of the matrix.

The second SiPM type was MPPC S10931-100P (Hamamatsu)~\cite{hamamatsu}: see Figure~\ref{image:fig_5x5el_Ham_3x3mm2},  Figure~\ref{image:fig_gain_CPTA_149-35__MPPC_S10931-100P__S13360-6050PE} and Figure~\ref{image:fig_noise_CPTA_149-35__MPPC_S10931-100P__S13360-6050PE}). At 87~K, it had a lower dark count rate (about 0.6~Hz/mm$^2$) and half as much maximum gain ($5\cdot10^5$) with respect to MRS APD 149-35~\cite{Bondar2015}. The 5$\times$5 SiPM matrix made from these SiPMs demonstrated stable operation for more than 20 cooling/heating cycles. However, this SiPM type has a narrow operating voltage range, resulting in substantial gain variations from channel to channel when powered by the same voltage.

The third (most successful) SiPM type was MPPC S13360-6050PE (Hamamatsu)~\cite{hamamatsu}: see Figure~\ref{image:fig_SiPM_11x11_matrix_photo},  Figure~\ref{image:fig_gain_CPTA_149-35__MPPC_S10931-100P__S13360-6050PE} and Figure~\ref{image:fig_noise_CPTA_149-35__MPPC_S10931-100P__S13360-6050PE}. At 87~K, it demonstrated a low dark count rate (about 0.1~Hz/mm$^2$) and high gains reaching $9\cdot10^6$ (these characteristics were measured following the procedure described in~\cite{Bondar2011, Bondar2015}). In addition, the MPPC S13360-6050P has a lower operating voltage and wider voltage range compared to MPPC S10931-100P, which significantly facilitated its use. The real matrix size was 11$\times$11 channels, of which only the central part of 5$\times$5 channels was active in the current measurements. This SiPM matrix demonstrated stable operation over 30 cooling/heating cycles and still is being used in our experimental setup.

\section{Experimental setup}

\begin{figure}[ht!]
	\center{\includegraphics[width=0.6\columnwidth]{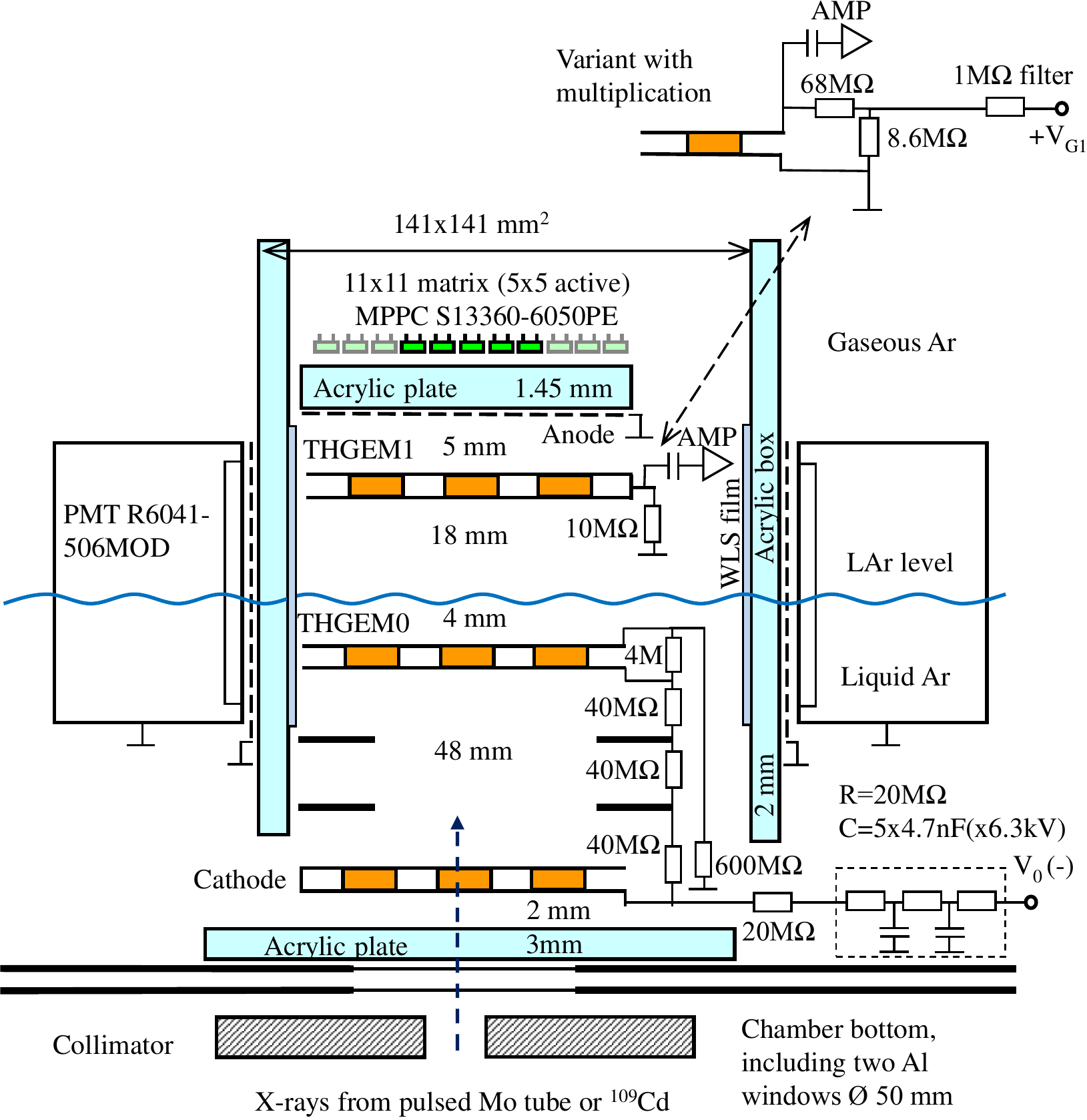}}
	\caption{Schematic view of the experimental setup. The electric fields lines in the TPC were presented elsewhere \cite{Bondar2019_field_sim}}
	\label{image:fig_3PMT_1PMT_THGEM}
\end{figure}

Figure~\ref{image:fig_3PMT_1PMT_THGEM} shows the experimental setup of the Novosibirsk group of the DarkSide collaboration. 
It included a 9-liter cryogenic chamber filled with 2.5 liters of liquid argon. The detector was operated in a two-phase mode in the equilibrium state at a saturated vapor pressure of 1.00~atm and temperature of 87.3~K.
Argon, of initial purity of 99.998 \%, was additionally purified from electronegative impurities during each cooling cycle by an Oxisorb filter, providing electron life-time in the liquid exceeding 100~$\mu$s~\cite{CRADELGap17}. 

The detector was a two-phase LAr time-projection chamber (TPC) composed of the drift (48~mm thick) and electron emission (4~mm thick) regions, in the liquid phase, and the EL gap (18~mm thick), in the gas phase.  
To form these regions, we used THGEM (Thick Gas Electron Multipliers,~\cite{Breskin2009}) electrodes instead of the more conventional wire grids, providing the advantage of electrode rigidity that allowed to avoid wire grid sagging. All electrodes had the same active area of 10$\times$10 cm$^2$. The THGEM geometrical parameters were the following: dielectric (FR-4) thickness of 0.4 mm, hole pitch of 0.9 mm, hole diameter of 0.5 mm and hole rim of 0.1 mm, optical transparency at normal incidence of 28\%.

The drift region was formed by a cathode electrode, field-shaping electrodes and THGEM0 (interface THGEM), immersed in the liquid layer. These were biased through a resistive high-voltage divider placed within the liquid. THGEM0 was biased in a way to provide a transmission of drifting electrons from the drift region to that of electron emission: the electrons drifted successively from a lower to higher electric field region. The electron transmission efficiency, defined by the voltage applied across THGEM0 and its geometrical parameters, was calculated in~\cite{Bondar2019_field_sim} to be 62\%.

THGEM1 was placed in the gas phase above the liquid and acted either as an anode of the EL gap (grounded through a resistor) or an electron multiplication element of the combined THGEM/SiPM-matrix multiplier (i.e. operated in electron avalanche mode), coupled to the EL gap.

The liquid level in the EL gap was monitored with an accuracy of 0.5~mm, being calculated from the amount of condensed Ar using a computer-aided design (CAD) software, the latter verified in special calibration runs with THGEM1 operated as a capacitive liquid level meter~\cite{Bondar2020}.

Three different readout configurations, corresponding to three EL mechanisms, were used in the measurements. In the first configuration, based on the ordinary EL mechanism, the EL gap was viewed by four compact cryogenic  PMTs R6041-506MOD~\cite{CryoPMT15}, located on the perimeter of the gap and electrically insulated from it by an acrylic box. Three of four PMTs were made sensitive to the VUV via WLS films (based on TPB in a polystyrene matrix~\cite{Gehman2013}) deposited on the inner box surface facing the EL gap, in front of these PMTs. We designate this configuration as 3PMT~+~WLS.

The second readout configuration corresponds to the concept of direct SiPM-matrix readout (see Figure~\ref{fig_conceptual_scheme}), based on the NBrS EL mechanism. Here the SiPM matrix, placed in the gas phase, is directly coupled to the EL gap, with light reaching the SiPMs through the holes of THGEM1 and the acrylic plate, the latter being used as additional electrical insulation. The SiPM matrix (see Figure~\ref{image:fig_SiPM_11x11_matrix_photo}) was made from MPPCs 13360-6050PE~\cite{hamamatsu} operated at an overvoltage of 5.6~V; their properties were described in the previous section. 
Taking into account the transmission of the acrylic plate in front of the matrix (see Figure~\ref{fig_reduced_EL_for_all_mech}), the SiPM matrix sensitivity ranges from the near UV (360~nm) to the NIR (1000~nm).
The contribution of crosstalk from the VUV, re-emitted by WLS on the side walls to the signal recorded by the SiPM matrix, was negligible as shown by experiments under similar conditions without WLS.

The third configuration corresponds to the concept of THGEM/SiPM-matrix readout (see Figure~\ref{fig_conceptual_scheme}), based on the avalanche scintillation mechanism. Here the combined THGEM/SiPM-matrix multiplier is coupled to the EL gap. In this case, a voltage is applied across THGEM1 (see top part of Figure~\ref{image:fig_3PMT_1PMT_THGEM}).
In addition to avalanche scintillations in the NIR, the SiPM matrix also recorded NBrS electroluminescence from the EL gap; its contribution however was negligible (of about 3\% at THGEM1 charge gain of 37).

It should be remarked that the detector was optimized for studying the all three readout techniques in the same experimental setup, rather than for reaching the maximum light yields. In particular for direct SiPM-matrix readout, the THGEM1 electrode acted as an optical mask, significantly (nine times) reducing the light flux: first, due to optical transparency  at normal incidence, of 28\%, and, second, due to angle dependence factor for optical transmission, of 40\% (determined by Monte-Carlo simulation). This, however, does not prevent in the following to assess the maximum light yields and detection thresholds that would be achieved under optimal conditions.

The detector was irradiated from outside either by X-rays from a pulsed X-ray tube with Mo anode, with an average deposited energy in liquid Ar of 25~keV~\cite{XRayYield16}, or by gamma rays from a $^{109}$Cd source~\cite{109Cd_2018}. 
To study the position resolution of the detector, a narrow beam of gamma-rays and X-rays was provided by a collimator with a hole diameter of 2~mm.

The signals from the PMTs were amplified using fast 10-fold amplifiers CAEN N979 and then re-amplified with linear amplifiers with a shaping time of 200~ns. The signals from 3PMT+WLS were summed (using CAEN N625 unit).
The signals from each SiPM were transmitted to amplifiers with a shaping time of 40~ns, via twisted pair wires. 
The charge signal from THGEM1 was recorded using a calibrated chain of a preamplifier and a shaping amplifier.
All amplifiers were placed outside the two-phase detector.

The SiPM signal amplitude was defined in terms of the number of recorded photoelectrons. 
The contribution of SiPM crosstalk (between the pixels) was accounted for and subtracted accordingly.
One of the channels of the SiPM matrix was inactive during data acquisition and the photoelectron number in it was determined as the average of two adjacent channels (see Figure~\ref{image:fig_171123_coorg_2Dgr_xray_2mm_MC_20kV}).

The DAQ system included both a 4-channel oscilloscope, model LeCroy WR HRO 66Zi, and a 64-channel Flash ADC CAEN V1740 (12~bits, 62.5~MHz): the signals were digitized and stored both in the oscilloscope and in a computer for further off-line analysis.
Other details of the experimental setup and measurement procedures can be found elsewhere~\cite{BondarINSTR2017EL,Buzulutskov2018}.

\section{EL gap yield for direct SiPM-matrix readout}

The performance of the two-phase detector with direct SiPM-matrix readout is characterized by the EL gap yield.
It is defined as the number of photoelectrons (PE) recorded by the SiPM matrix in total per drifting electron in the EL gap.

To measure the EL gap yield, a $^{109}$Cd gamma-ray source was used.
The emission spectrum of this source includes low-energy (22-25~keV) and high-energy lines: namely the characteristic lines of W (59~keV), which was used as a radionuclide substrate, and the 88~keV line of $^{109}$Cd itself~\cite{109Cd_2018}.
Due to insufficient energy resolution, the 59 and 88~keV lines could not be separated; therefore their weighted average energy (82~keV~\cite{109Cd_2018}) was used in the analysis.

\begin{figure}[ht!]
	\center{\includegraphics[width=0.6\columnwidth]{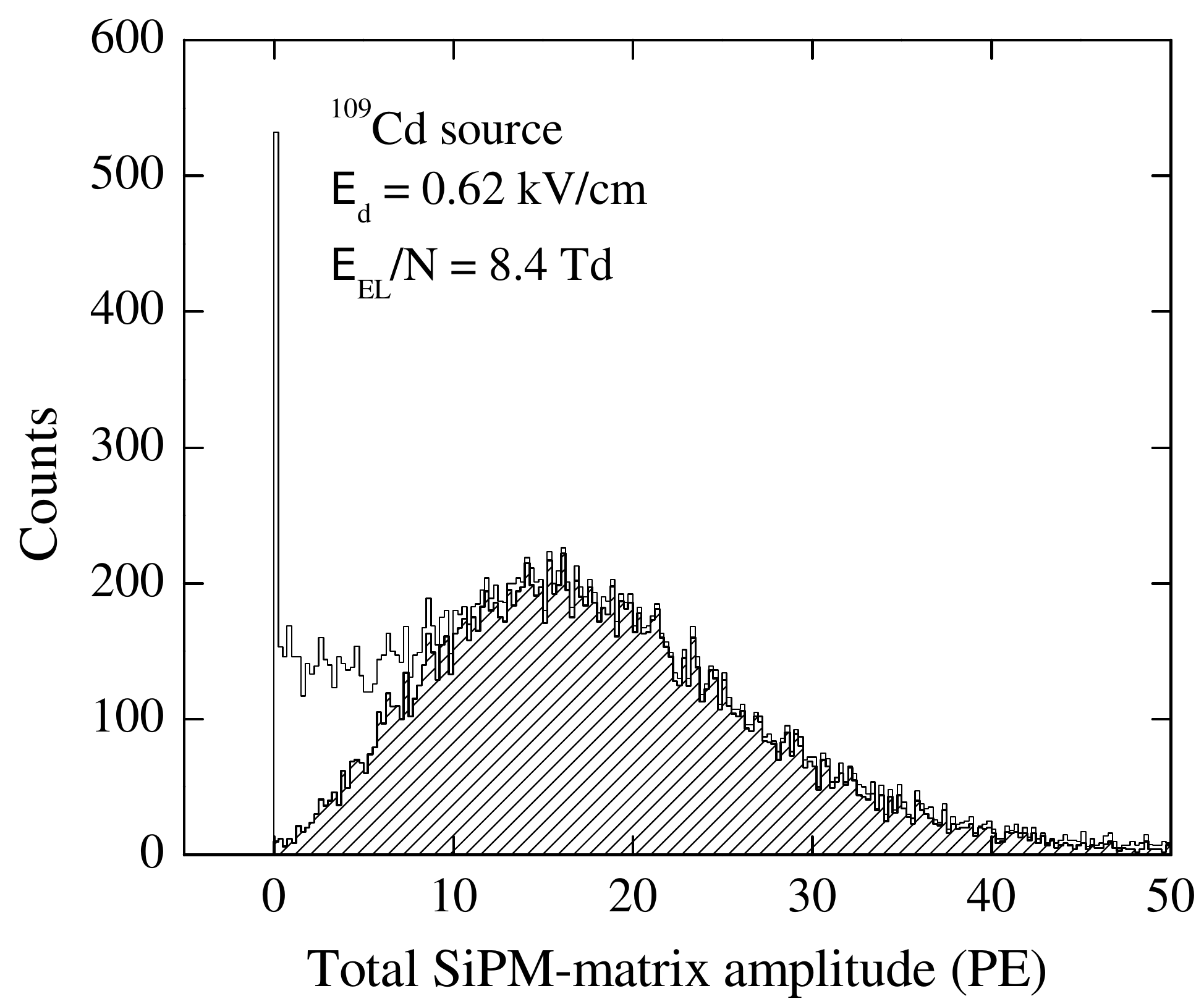}}
	\caption{Direct SiPM-matrix readout: amplitude spectrum of the total SiPM-matrix signal obtained with $^{109}$Cd source. The hatched area corresponds to the higher energy peak of the 3PMT+WLS signals (see Figure~\ref{image:fig_171123_3PMT_Cd_and_Bkg_Npe_spec_6mm_20kV})}
	\label{image:fig_171123_SiPM_48V_Cd_Npe_spec_82keVcuts_6mm_20kV}
\end{figure}

\begin{figure}[ht!]
	\center{\includegraphics[width=0.6\columnwidth]{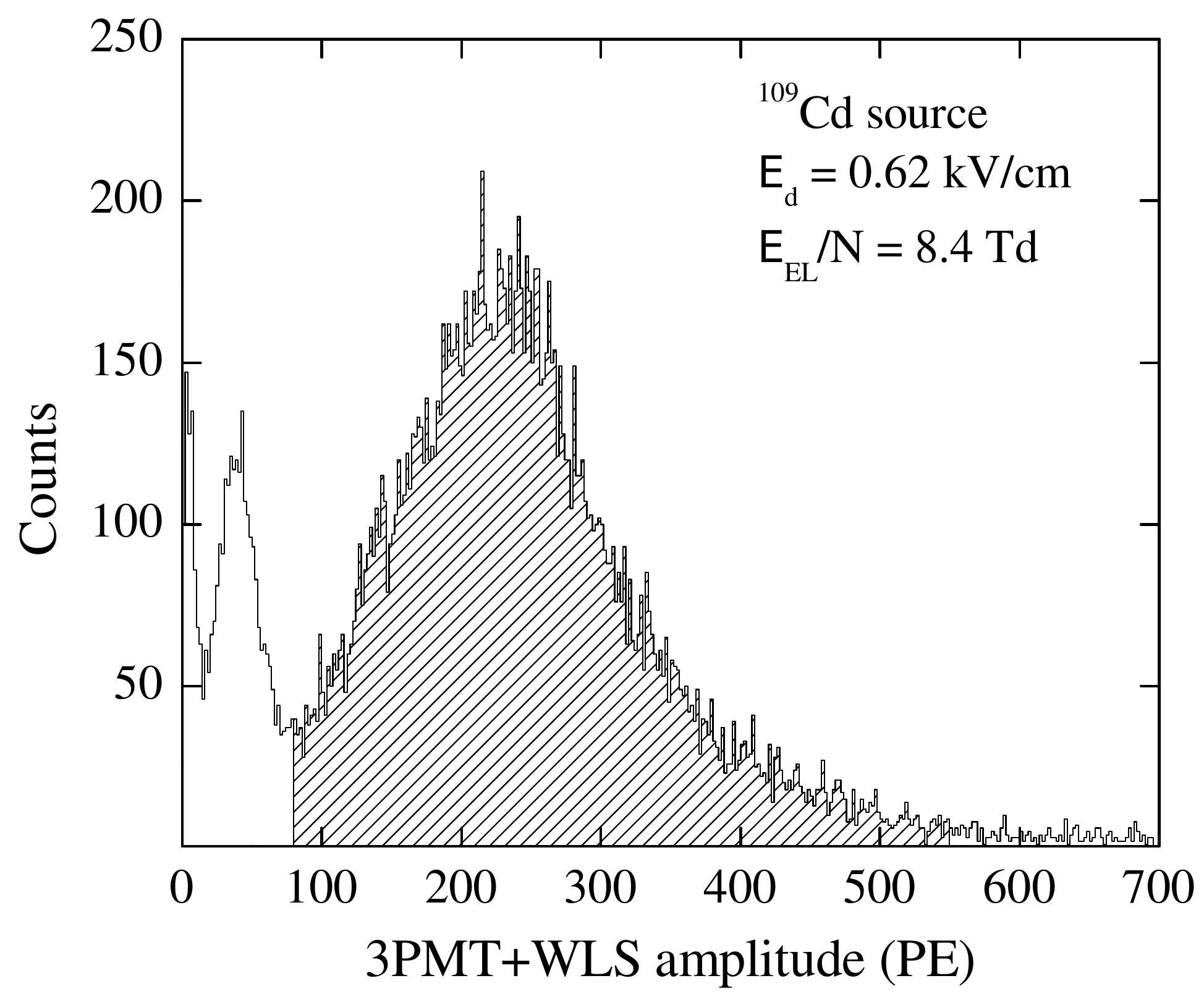}}
	\caption{Direct SiPM-matrix readout: amplitude spectrum of the total 3PMT+WLS signal obtained with $^{109}$Cd source for the maximum field in the EL gap. The higher energy peak, corresponding to 59-88~keV gamma-rays, is hatched}
	\label{image:fig_171123_3PMT_Cd_and_Bkg_Npe_spec_6mm_20kV}
\end{figure}

Due to the small photoelectron number, it was not possible to directly separate the low and high energy parts in the SiPM amplitude spectrum: see Figure~\ref{image:fig_171123_SiPM_48V_Cd_Npe_spec_82keVcuts_6mm_20kV}.  On the other hand, the 3PMT+WLS amplitude was high enough to make such a separation: see Figure~\ref{image:fig_171123_3PMT_Cd_and_Bkg_Npe_spec_6mm_20kV}.

\begin{figure}[ht!]
	\center{\includegraphics[width=0.65\columnwidth]{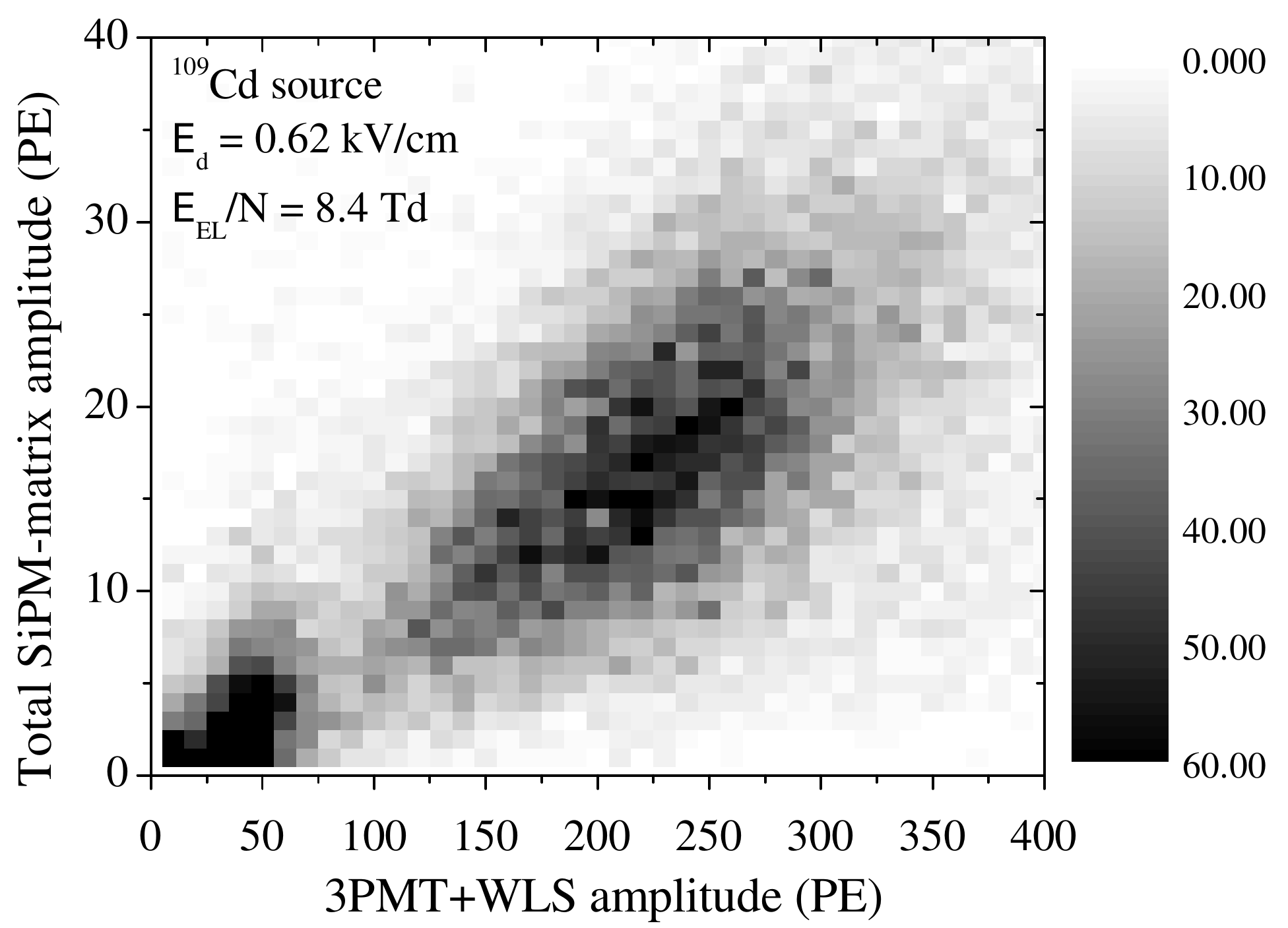}}
	\caption{Direct SiPM-matrix readout: correlation between the amplitude of the total SiPM-matrix and 3PMT+WLS signals}
	\label{image:fig_PMTSiPMCorrel_Cd_20kV_PMT750_12dB_coll_6mm_real}
\end{figure}

Since the 3PMT+WLS and SiPM-matrix signals are correlated (see Figure~\ref{image:fig_PMTSiPMCorrel_Cd_20kV_PMT750_12dB_coll_6mm_real}), it is possible to separate the events with higher and lower energy in the SiPM-matrix amplitude spectrum, selecting appropriately the events in the 3PMT+WLS amplitude spectrum. This is seen in 
Figure~\ref{image:fig_171123_SiPM_48V_Cd_Npe_spec_82keVcuts_6mm_20kV} showing the SiPM-matrix amplitude spectrum, where the hatched area is obtained by selecting the 3PMT+WLS signals from the higher energy peak: see Fig~\ref{image:fig_171123_3PMT_Cd_and_Bkg_Npe_spec_6mm_20kV}.
Only the average photoelectron number of this (high-energy) part of the spectrum was used to determine the EL gap yield.

\begin{figure}[ht!]
	\centering
	\subfloat{\includegraphics[width=0.65\columnwidth]{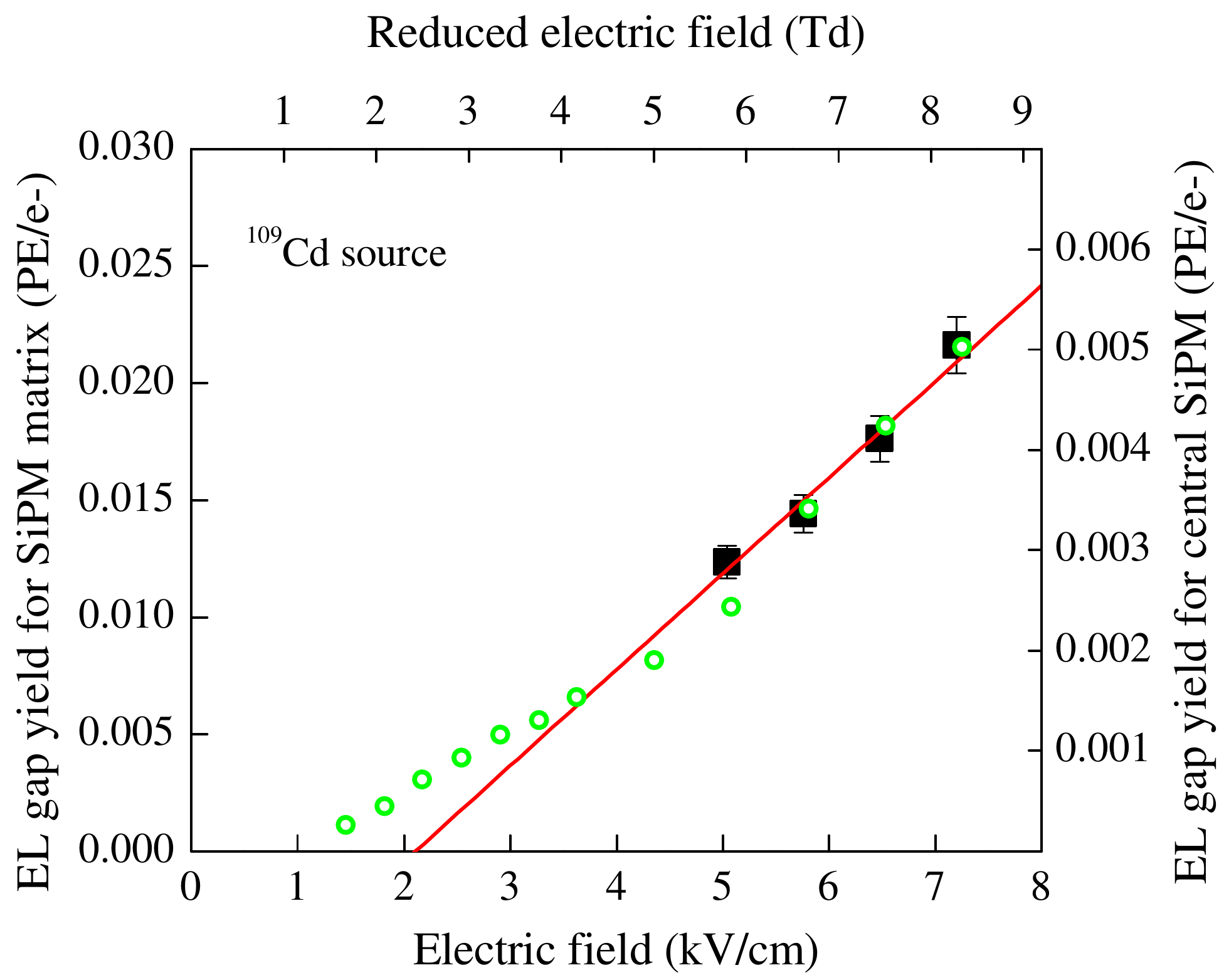}} 
	\caption[]{Direct SiPM-matrix readout: EL gap yield for the SiPM matrix in total (solid data points, left scale) as a function of the electric field or reduced electric field in the EL gap, at the average energy of 82~keV deposited by gamma-rays from $^{109}$Cd source in liquid Ar. The red line is a linear fit to the data points. For comparison, the EL gap yield for the central SiPM measured elsewhere~\cite{Bondar2019} is shown (open data points, right scale) \label{image:fig_SiPM_matrix_yield_48V_pe_e}}
\end{figure}

In addition, to calculate the EL gap yield, one has to know the charge emitted from the liquid into the EL gap. Since it was too small for direct recording (about 800~e$^-$), it was calculated theoretically using the data on ionization yields for electron recoils in liquid argon~\cite{XRayYield16} and on electron transmission through the THGEM0 electrode~\cite{Bondar2019_field_sim}.

The EL gap yield was obtained by dividing the average photoelectron number recorded by the SiPM matrix in total to the calculated charge. The EL gap yield obtained this way, as a function of the electric field in the EL gap,  is shown in Fig~\ref{image:fig_SiPM_matrix_yield_48V_pe_e}. At higher fields, between 5 and 8~Td, the field dependence is well described by a linearly growing function. For comparison, the EL gap yield for the central SiPM only, measured for wider field range in our previous work~\cite{Bondar2019}, is shown. One can see a good reproducibility of the linear field dependence in data overlap.

The maximum EL gap yield amounted to 0.022~PE/e$^-$ at an electric field in the EL gap of 7.3~kV/cm (corresponding to the reduced field of 8.4~Td), which corresponds to 0.2~PE per keV of the energy deposited in liquid Ar.
This value is fairly small. We will see in the following (section~\ref{minimum_detection_threshold}) that it can be significantly increased, by about an order of magnitude, for the optimal detector structure.

\section{THGEM/SiPM-matrix yield}\label{combined_multiplier}

Similarly to the EL gap yield with direct SiPM-matrix readout, we can define the yield of the combined THGEM/SiPM-matrix multiplier coupled to the EL gap (or THGEM/SiPM-matrix yield for short), as the number of photoelectrons recorded by the SiPM matrix per drifting electron in the EL gap.

\begin{figure}[ht!]
	\center{\includegraphics[width=0.55\columnwidth]{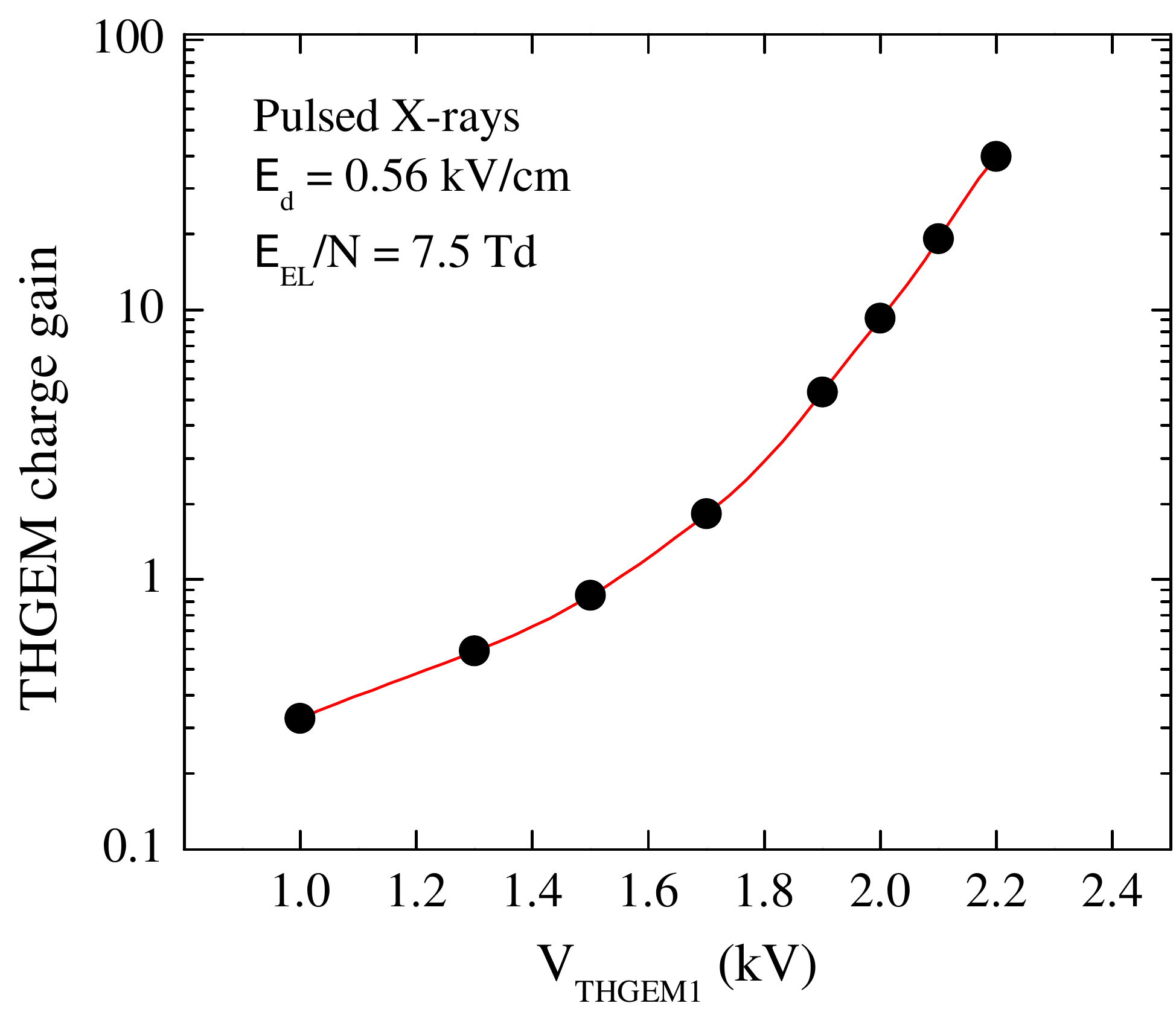}}
	\caption{THGEM/SiPM-matrix readout: charge gain of the THGEM1 multiplier as a function of the voltage across it, at fixed electric fields in the drift and EL regions}
	\label{image:fig_THGEM_Gain}
\end{figure}

Here, THGEM1 was operated in electron avalanche mode, its charge gain being measured using a pulsed X-ray tube (similarly to~\cite{CRAD_with_THGEM_2013}). 
Figure~\ref{image:fig_THGEM_Gain} shows THGEM1 charge gain as a function of the voltage applied across it, at fixed drift and EL gap electric fields.

\begin{figure}[ht!]
	\center{\includegraphics[width=0.55\columnwidth]{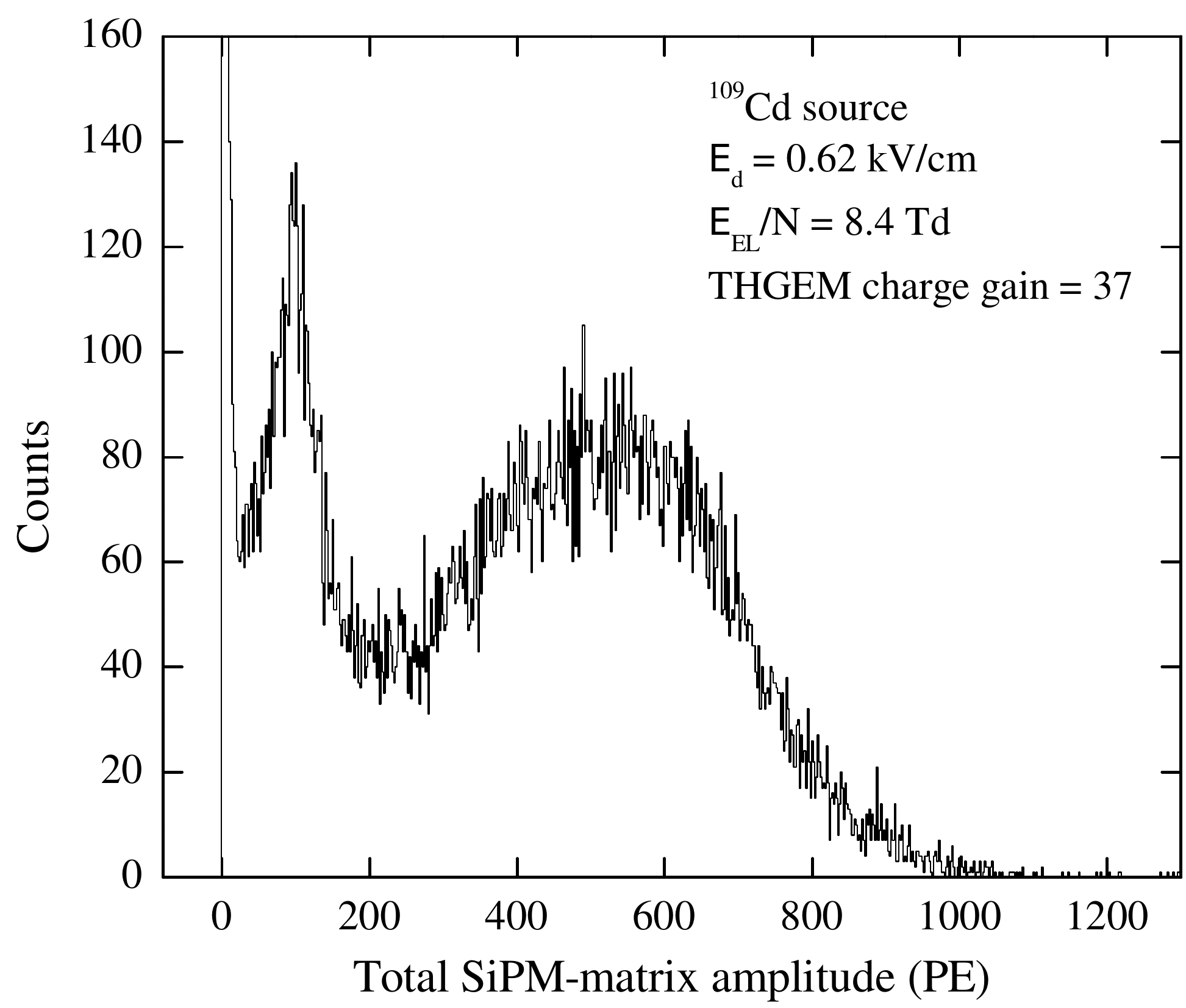}}
	\caption{THGEM/SiPM-matrix readout: amplitude spectrum of the total SiPM-matrix signal obtained with $^{109}$Cd source, at THGEM1 charge gain of 37}
	\label{image:fig_171130_SiPM_48V_Cd_Npe_spec_2mm_20kV_gain37}
\end{figure}

The yield of the combined THGEM/SiPM-matrix multiplier was measured at two THGEM1 voltages, of 2.0 and 2.2~kV, corresponding to THGEM1 charge gain of 9 and 37. Using the $^{109}$Cd source, the amplitude spectra of the signals from the SiPM matrix were recorded.
Due to sufficient energy resolution, it was possible to separate the low-energy (22-25~keV) and high-energy events (59-88~keV) without using the 3PMT+WLS signals: see Figure~\ref{image:fig_171130_SiPM_48V_Cd_Npe_spec_2mm_20kV_gain37}.
It should be remarked that the degradation of energy resolution of the combined THGEM/SiPM-matrix multiplier due to inherent avalanche fluctuations, compared to the direct SiPM-matrix readout, becomes insignificant at higher statistics of drifting electrons (exceeding 10) due to multiple spectrum convolution.

\begin{figure}[ht!]
	\centering
	\subfloat{\includegraphics[width=0.6\columnwidth]{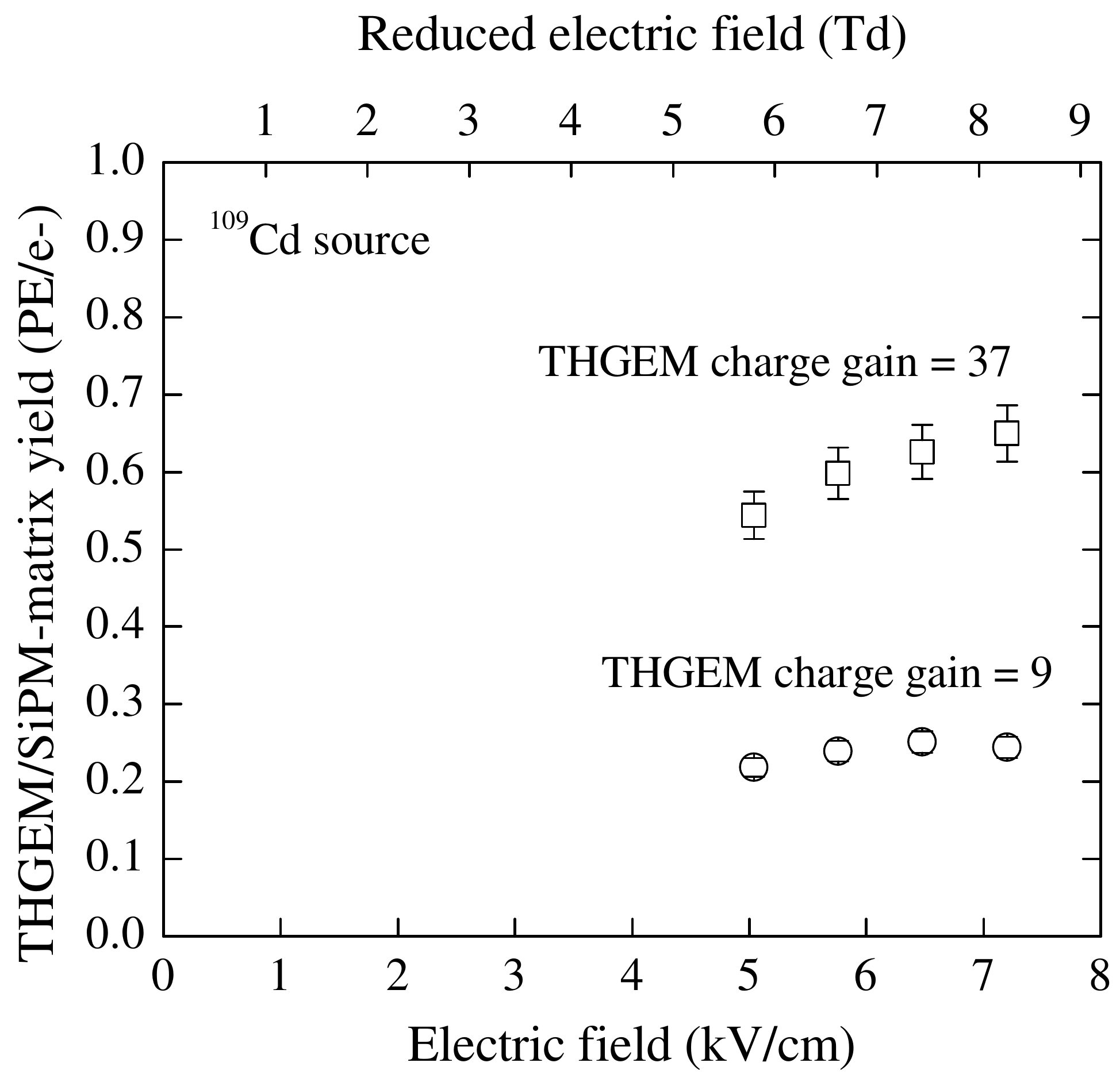}}  
	\caption[]{THGEM/SiPM-matrix readout: THGEM/SiPM-matrix yield as a function of the electric field or reduced electric field in the EL gap at the average energy of 82~keV, deposited by gamma-rays from $^{109}$Cd source in liquid argon, measured at two THGEM charge gains\label{image:fig_THGEM_SiPM_matrix_yield_48V_pe_e}}	
\end{figure}

Similarly to direct SiPM-matrix readout, the average number of photoelectrons for the high-energy part of the spectrum was defined and then divided by the calculated charge emitted into the EL gap. The THGEM/SiPM-matrix yield obtained this way is shown in Figure~\ref{image:fig_THGEM_SiPM_matrix_yield_48V_pe_e}. One can see that the yield is sensitive to the THGEM gain, rather than to the electric field in the EL gap. This is because the THGEM/SiPM-matrix yield, being first of all proportional to the THGEM charge gain, weakly depends on the external electric field.

The maximum THGEM/SiPM-matrix yield amounted to 0.65~PE/e$^-$ at a charge gain of 37 and electric field in the EL gap of 7.3~kV/cm, which corresponds to 6.2~PE per keV of the energy deposited in liquid Ar. One can see that even at such a  moderate THGEM gain, the amplitude yield of the THGEM/SiPM-matrix readout is considerably (by more than order of magnitude) increased compared to the direct SiPM-matrix readout. In section~\ref{minimum_detection_threshold}, we will estimate the detection thresholds for nuclear recoils for these readout techniques, under the optimal conditions.

\section{$x$, $y$ coordinate reconstruction algorithm} \label{spatial_resolution}

One of the main advantages of the SiPM matrix readout is the high reconstruction accuracy of the event coordinates in $x$, $y$ plane of the two-phase detector. In this section and the next one, the reconstruction algorithm and the position resolution will be described. These results were obtained in the two-phase detector when irradiated by a pulsed X-ray tube or $^{109}$Cd source through a 2~mm collimator.

Let us define the following values: $x_0$ is the true coordinate of the X-ray photon interaction point in the liquid, $X_i$ is the coordinate of the center of the $i$-th element of the SiPM matrix, $N_{i}$ is the number of photoelectrons recorded by the $i$-th element of the SiPM matrix, $N_{ch}$ is the number of channels of the SiPM matrix, $x_{exp}$ and $x_{sim}$ are coordinates of the interaction point reconstructed from experimental data and simulation, respectively.

The center of gravity (CoG) algorithm is one of the simplest methods widely used for coordinate reconstruction~\cite{Landi2002}.
According to this algorithm, $x_{exp}$ is calculated using the following formula:

\begin{equation}
\label{eq.CoG_def} x_{exp} = \left( \sum_{i=1}^{N_{ch}} X_{i} \cdot N_{i} \right) / \left( \sum_{i=1}^{N_{ch}} N_{i} \right) \, .
\end{equation}
Similar formulas are used for $y$ coordinate.

A well-known feature of the CoG algorithm is the compression effect, resulting in that the reconstructed coordinates are biased to the center of the matrix~\cite{Fabbri2013}.
To eliminate such a systematic bias, it is necessary to find the dependence of the reconstructed coordinate on the true one: $x_{exp}(x_{0})$.
Since in our case $x_{0}$ is not known from experimental data, the desired dependence $x_{exp}(x_0)$ is determined by simulation: $ x_{sim}(x_0)$ and its inverse function $x_{0}(x_{sim})$~\cite{Bondar2018_THGEM_GAPD}.

To find these dependences, it is obviously necessary to know how detected photons (i.e. photoelectrons) are distributed over the elements of the SiPM matrix for the given coordinates of the interaction point ($x_0$, $y_0$).
This distribution over the elements of the SiPM matrix ($N_i$) is described by the following expression:
\begin{equation} \label{eq.PE_LRF}
N_{i}(x_0, y_0) = N_0 \cdot LRF_{i}(X_i - x_0, Y_i - y_0) \, ,
\end{equation}
where $N_0$ is the number of photons emitted at the interaction point ($x_0$, $y_0$), and $LRF_i$ is the so-called Light Response Function~\cite{Solovov2012}, i.e. the fraction of photons registered by $i$-th element of the SiPM matrix for a given interaction point ($x_0$, $y_0$).
It is obvious that $LRF_i$ has a maximum when $X_i - x_0 = Y_i - y_0 = 0$, i.e. when the projection of the interaction point is in the center of the channel.

In principle, $LRF_i$ can be calculated by Monte Carlo (MC), simulating the propagation of photons in the detector.
However, this is a difficult task, since the correct description of the properties of all optical surfaces is not always achievable.
In this regard, $LRF_i$ is determined empirically, from experimental data.

\begin{figure}[ht!]
	\center{\includegraphics[width=0.6\columnwidth]{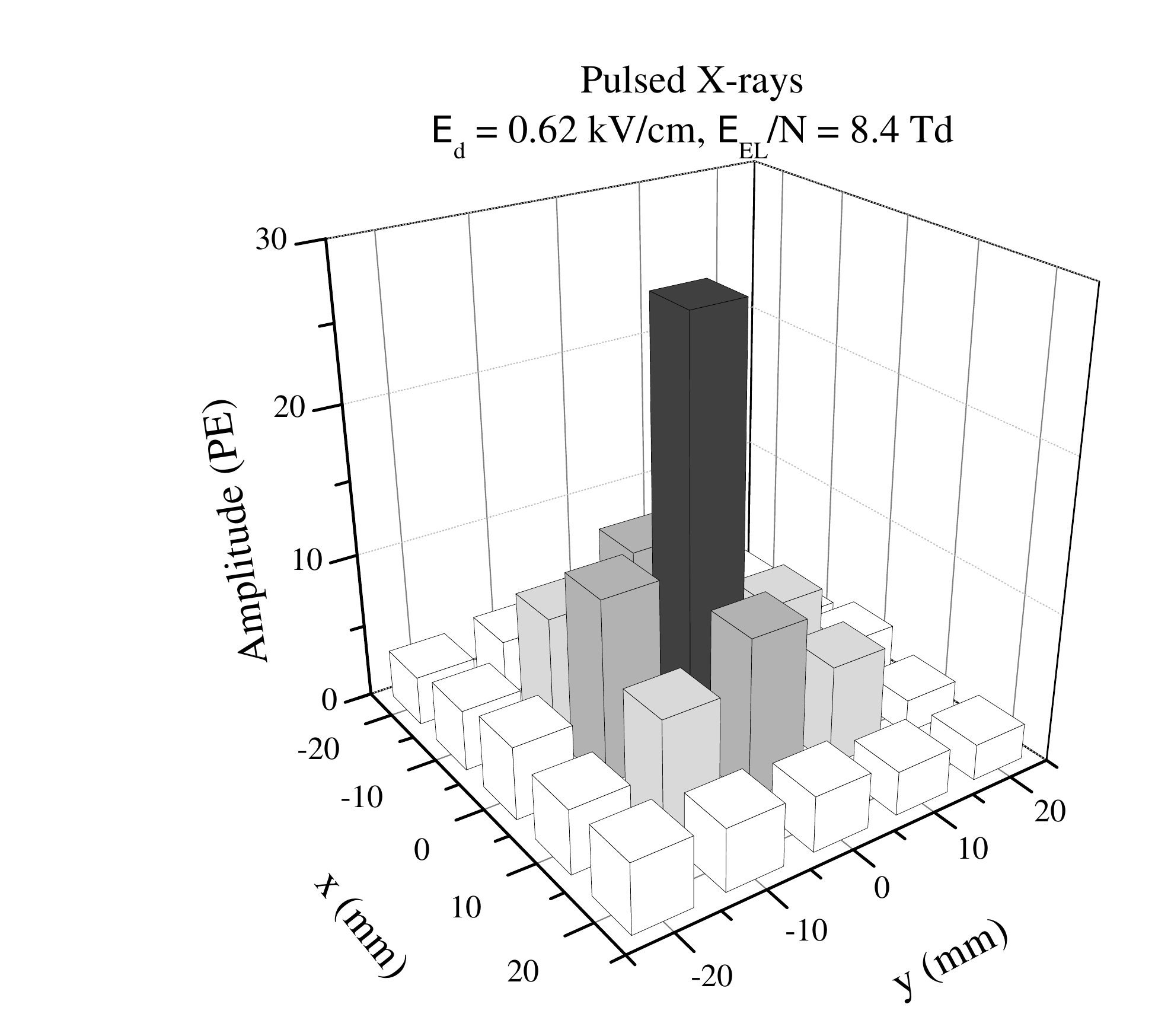}}
	\caption[]{Direct SiPM-matrix readout: the averaged distribution of photoelectrons over the SiPM-matrix channels in x,y plane for ``central'' events, in which the distribution's maximum is at the central channel. The data were obtained at the maximum reduced EL field, of 8.4~Td, when the detector was irradiated by pulsed X-rays}
	\label{image:fig_171123_PE_distr_xray_2mm_20kV}
\end{figure}

\begin{figure}[ht!]
	\centering
	\subfloat{\includegraphics[width=0.45\columnwidth]{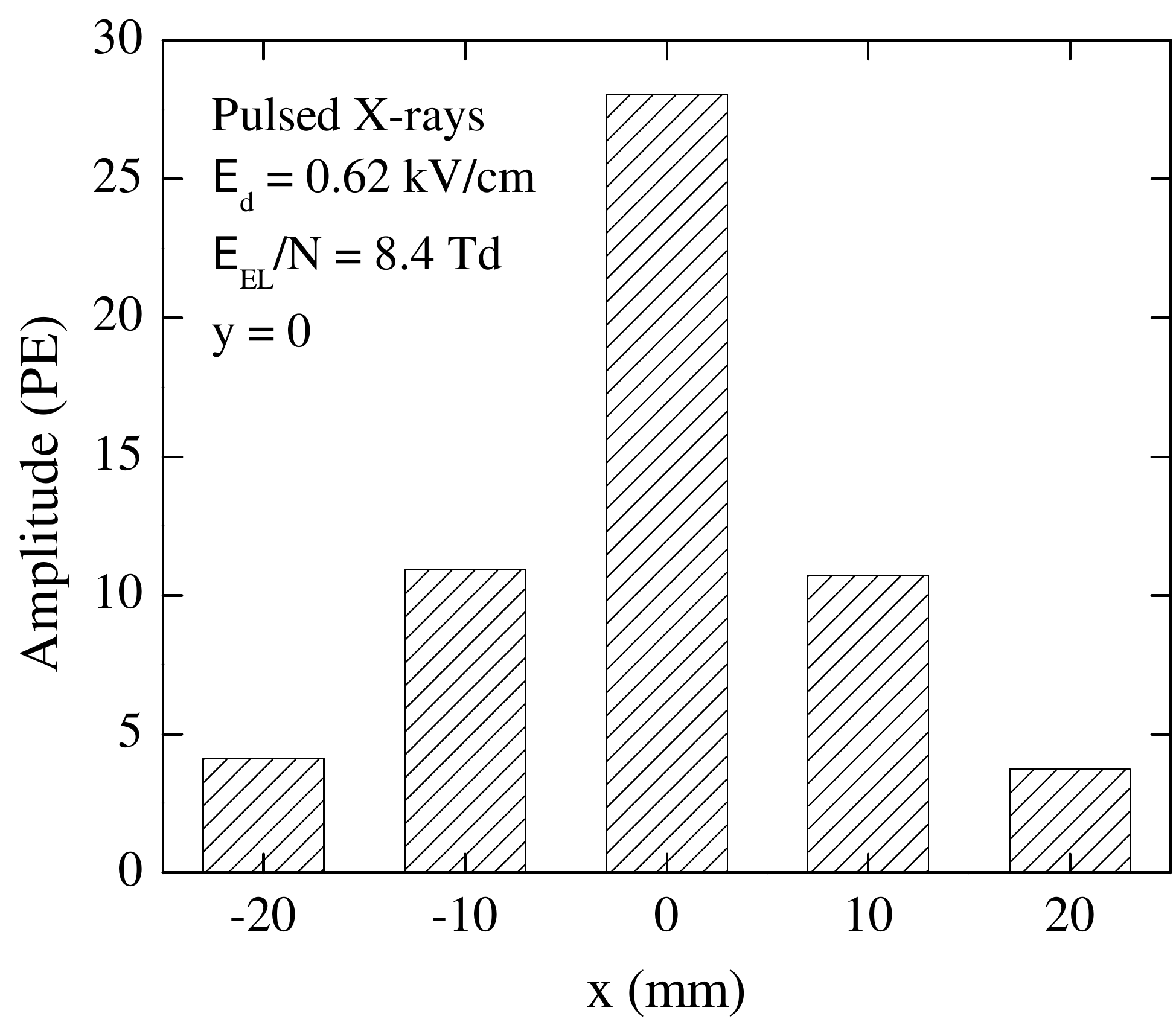}} \hspace{5mm}
	\subfloat{\includegraphics[width=0.45\columnwidth]{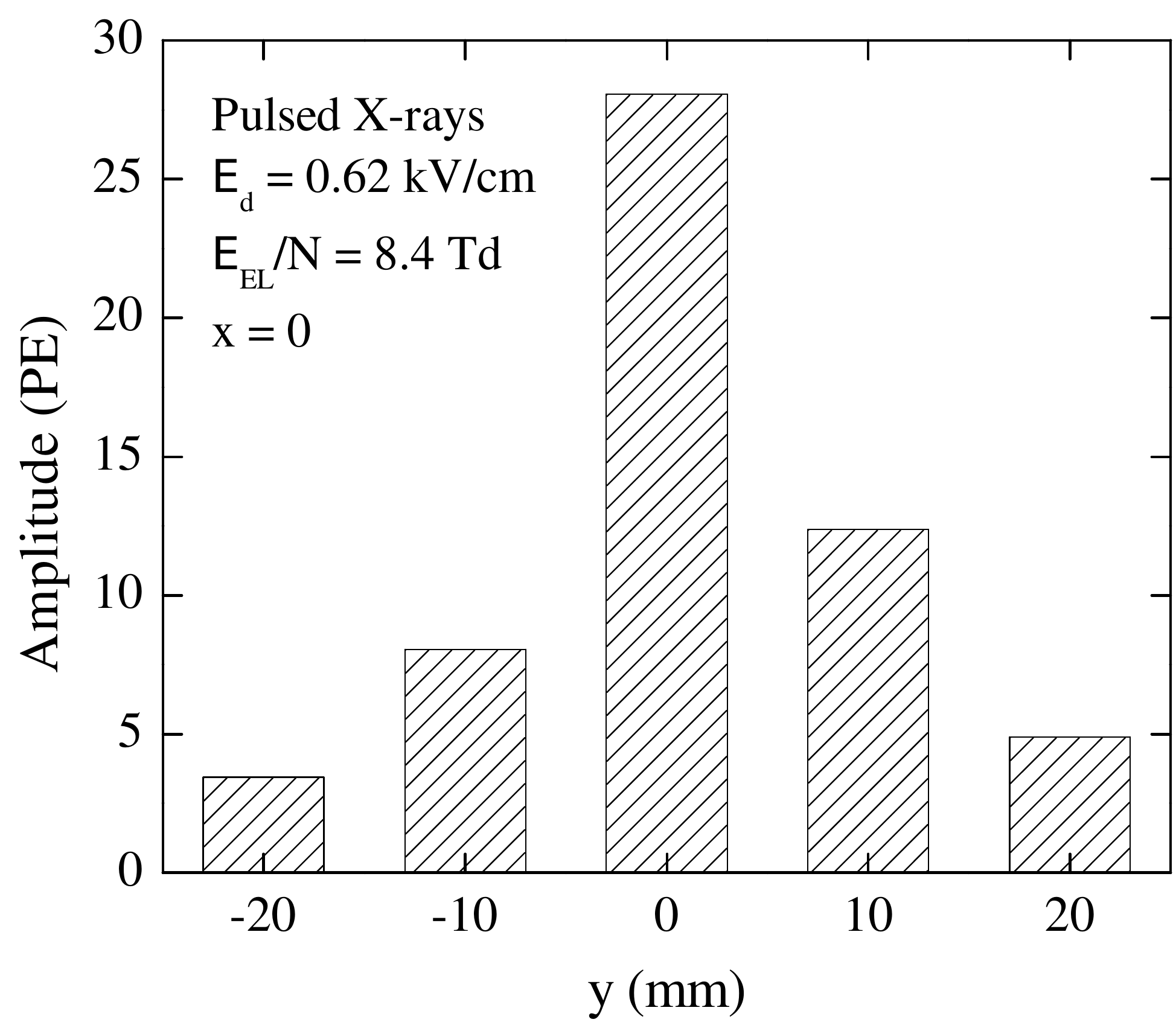}} 
	\caption[] {\label{image:fig_PE_XYProjection_x_ray_20kV_PMT550_0dB_coll_2mm} 
		Direct SiPM-matrix readout: 2D cross-section of Figure~\ref{image:fig_171123_PE_distr_xray_2mm_20kV} at $y$ = 0 (left) and $x$ = 0 (right)}
\end{figure}

To determine $LRF_i$, first of all, an averaged distribution of photoelectrons $N_{i}$ over the channels of the SiPM matrix for ``central'' events (for which the maximum of the distribution hits the central channel) was obtained.
Figure~\ref{image:fig_171123_PE_distr_xray_2mm_20kV} shows 3D distribution and Figure~\ref{image:fig_PE_XYProjection_x_ray_20kV_PMT550_0dB_coll_2mm} its 2D cross-sections for such ``central'' events. 

Next, we use the approximation that the  $LRF_i$ shape is the same for all SiPM-matrix channels. This approximation is justified by the fact that the interaction region in $x$,~$y$ plane was much smaller (less than 0.5~cm in diameter) than the active region of the detector ($10\times10$~cm$^2$). Thus, the $LRF$ obtained for the central channel could be used for all other SiPM-matrix channels.
The $LRF$ (up to scaling factor) was found from Figure~\ref{image:fig_171123_PE_distr_xray_2mm_20kV} using a linear interpolation.

\begin{figure}[ht!]
	\center{\includegraphics[width=0.6\columnwidth]{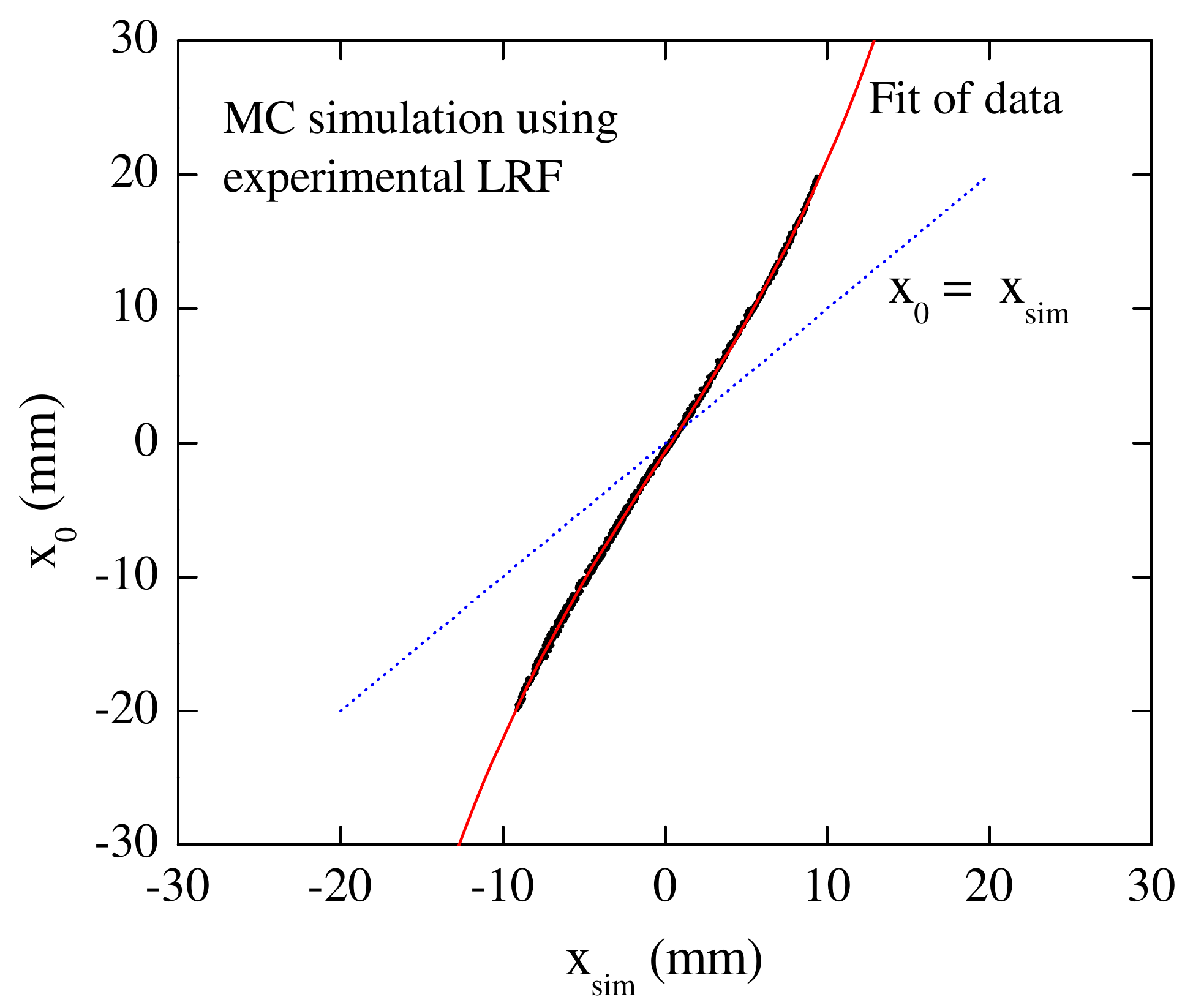}}
	\caption{Direct SiPM-matrix readout: true coordinate of interaction point $x_{0}$ as a function of reconstructed coordinate $x_{sim}$ (black dots), the latter obtained using MC simulation with the experimental $LRF$, and a fit of this dependence by polynomial function (red curve). For comparison, the dependence $x_0 = x_{sim}$ is shown (blue dotted line)
	}
	\label{image:fig_171123_MC_corrections_xray_2mm_20kV}
\end{figure}

Using the $LRF$ obtained this way, $x_0(x_{sim})$ and $y_0(y_{sim})$ dependencies were found.
To this end, $x_0$ and $y_0$ coordinates were randomly and uniformly generated in the range of ($-20$~mm, $20$~mm), and then $x_{sim}$ and $y_{sim}$ coordinates were calculated using the CoG formula~\eqref{eq.CoG_def}. Figure~\ref{image:fig_171123_MC_corrections_xray_2mm_20kV} shows the resulting $x_{0}(x_{sim})$ dependence obtained this way, along with the trivial $x_0=x_{sim}$ dependence (i.e. in the absence of systematic bias).

\section{Results of $x$, $y$ reconstruction}

\begin{figure}[ht!]
	\center{\includegraphics[width=0.6\columnwidth]{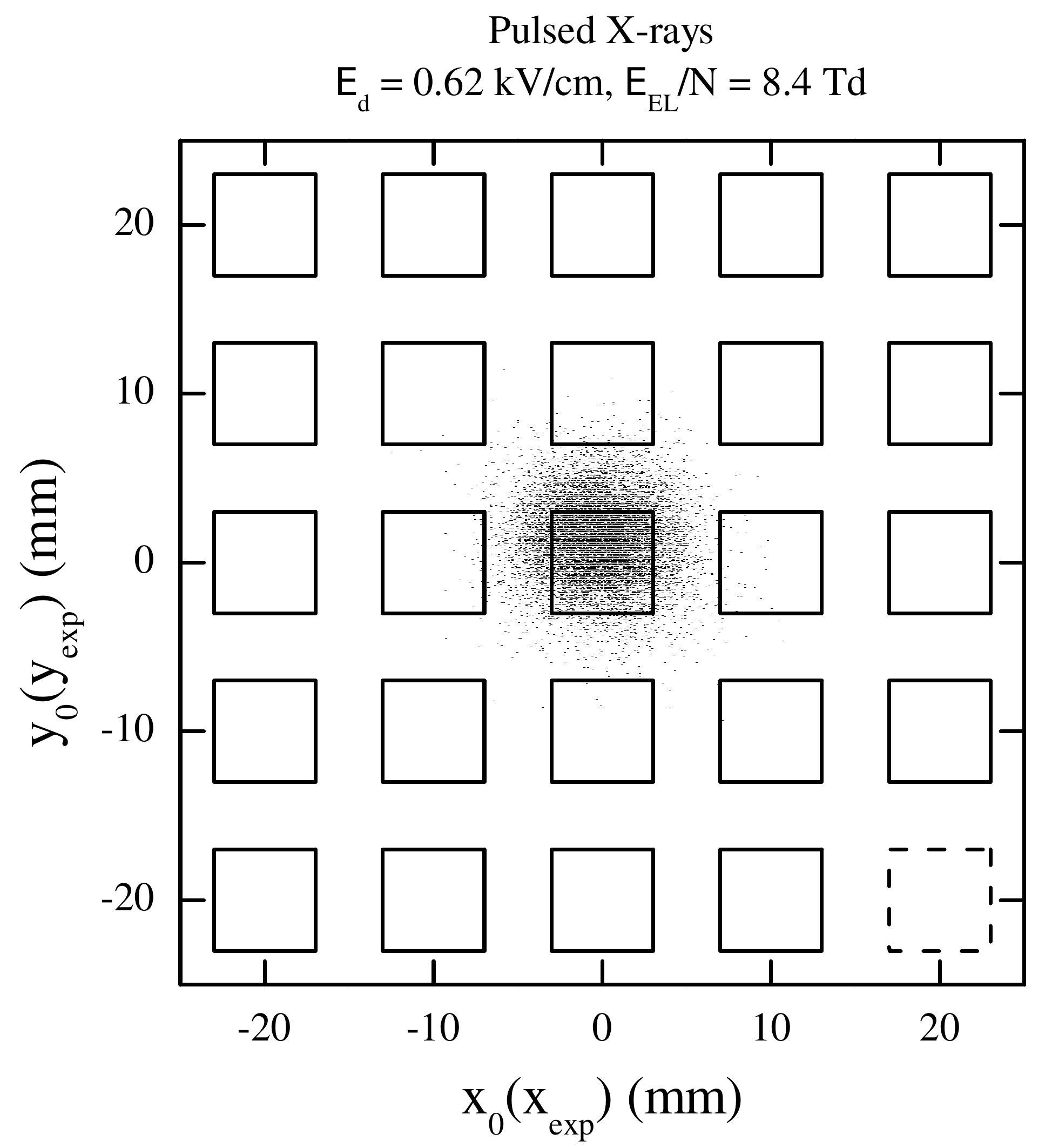}}
	\caption{ Direct SiPM-matrix readout: 2D coordinate distribution of reconstructed events in $x_0(x_{exp})$, $y_0(y_{exp})$ plane. The solid boxes are the active SiPMs and the dashed box is the inactive SiPM (the photoelectron number in which was determined as the average of two adjacent channels). The data were obtained at the maximum reduced EL field, of 8.4~Td, when the detector was irradiated by pulsed X-rays through a 2~mm collimator}
	\label{image:fig_171123_coorg_2Dgr_xray_2mm_MC_20kV}
\end{figure}

Applying the CoG algorithm to the experimental data and taking into account the corrections for a systematic bias using the fitted curve in Figure~\ref{image:fig_171123_MC_corrections_xray_2mm_20kV}, the desired event distributions over $x_0(x_{exp})$ and $y_0(y_{exp})$ were obtained. In particular, Figure~\ref{image:fig_171123_coorg_2Dgr_xray_2mm_MC_20kV} shows 2D distribution of the event coordinates in $x_0(x_{exp})$, $y_0(y_{exp})$ plane for direct SiPM-matrix readout, when the detector was irradiated by pulsed X-rays through a 2~mm collimator.

\begin{figure}[ht!]
	\centering
	\subfloat{\includegraphics[width=0.45\columnwidth]{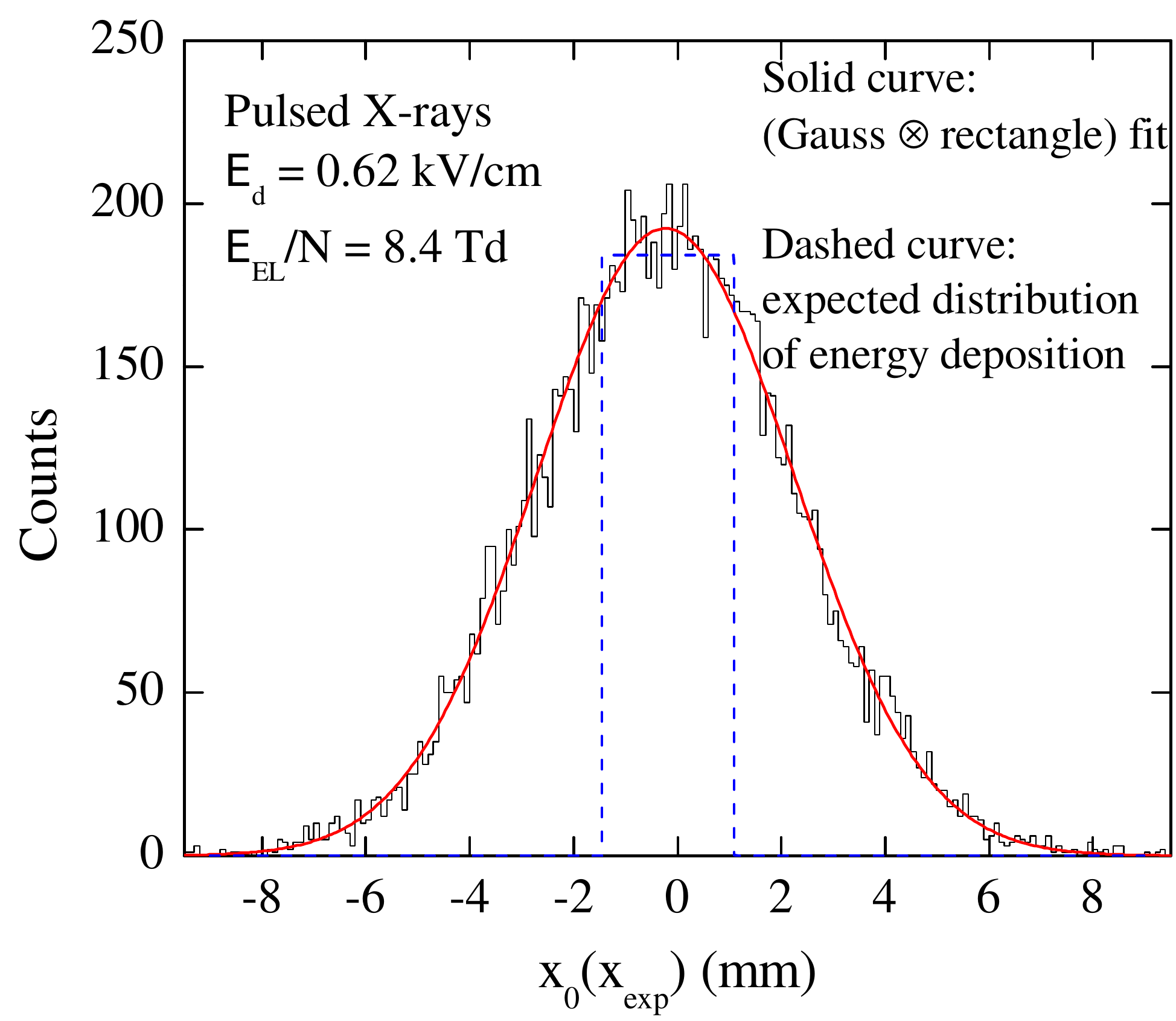}} \hspace{5mm}
	\subfloat{\includegraphics[width=0.45\columnwidth]{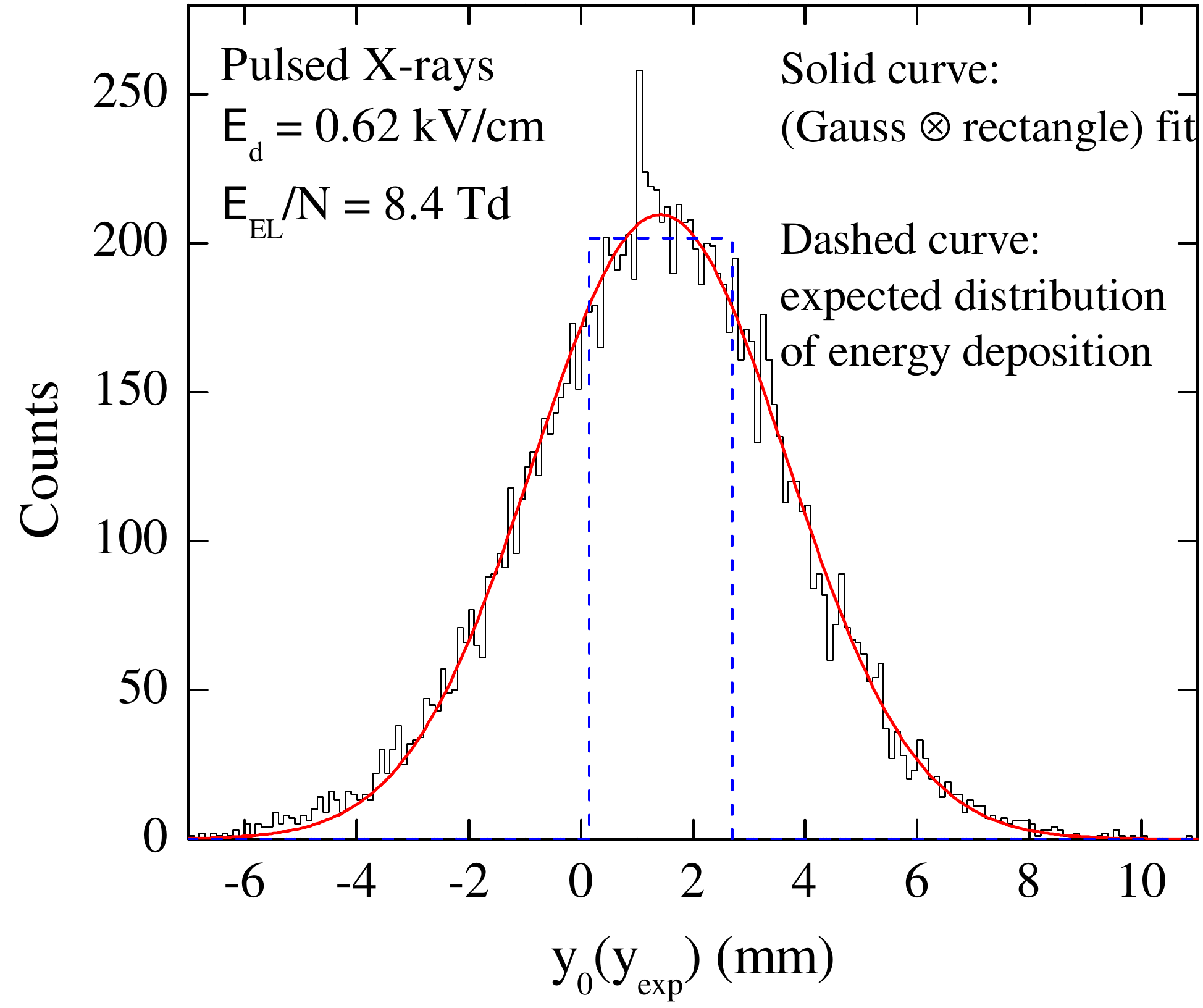}} 
	\caption[x,y distributions using CoG algorithm with corrections] {\label{image:fig_171123_XYcoorg_xray_2mm_MC_20kV} 
		Direct SiPM-matrix readout: projections of the 2D distribution presented in Figure~\ref{image:fig_171123_coorg_2Dgr_xray_2mm_MC_20kV} on $x$ and $y$ axes. Also shown are the fit of the $x_0(x_{exp})$ and $y_0(y_{exp})$ distributions (red curves) and the expected distributions of the true coordinates of the interaction region (blue dotted curves), defined by the positions of the X-ray tube and collimator and by the X-ray range in liquid Ar. Note that here the signal is produced by several X-ray photons in a pulse, with the average energy of 25 keV, absorbed in a thin (3 mm) liquid Ar layer near the cathode}
\end{figure}

Figure~\ref{image:fig_171123_XYcoorg_xray_2mm_MC_20kV} shows the projections of  Figure~\ref{image:fig_171123_coorg_2Dgr_xray_2mm_MC_20kV} on $x$ and $y$ axes.
The fit of the distribution on $x_0(x_{exp})$ and $y_0(y_{exp})$ (red curve) and the rectangular distribution of the true coordinate of the interaction point (blue dotted curve) are also shown.
The latter was determined geometrically taking into account the relative position of the radiation source and collimator and the X-ray range in liquid Ar.   
The fit function represented a convolution of this rectangular distribution with a Gaussian function.
The latter is defined by the detector resolution.
Thus, the fitting parameter of the Gauss function ($\sigma$) characterizes the position resolution of the detector.


\begin{figure}[ht!]
	\center{\includegraphics[width=0.6\columnwidth]{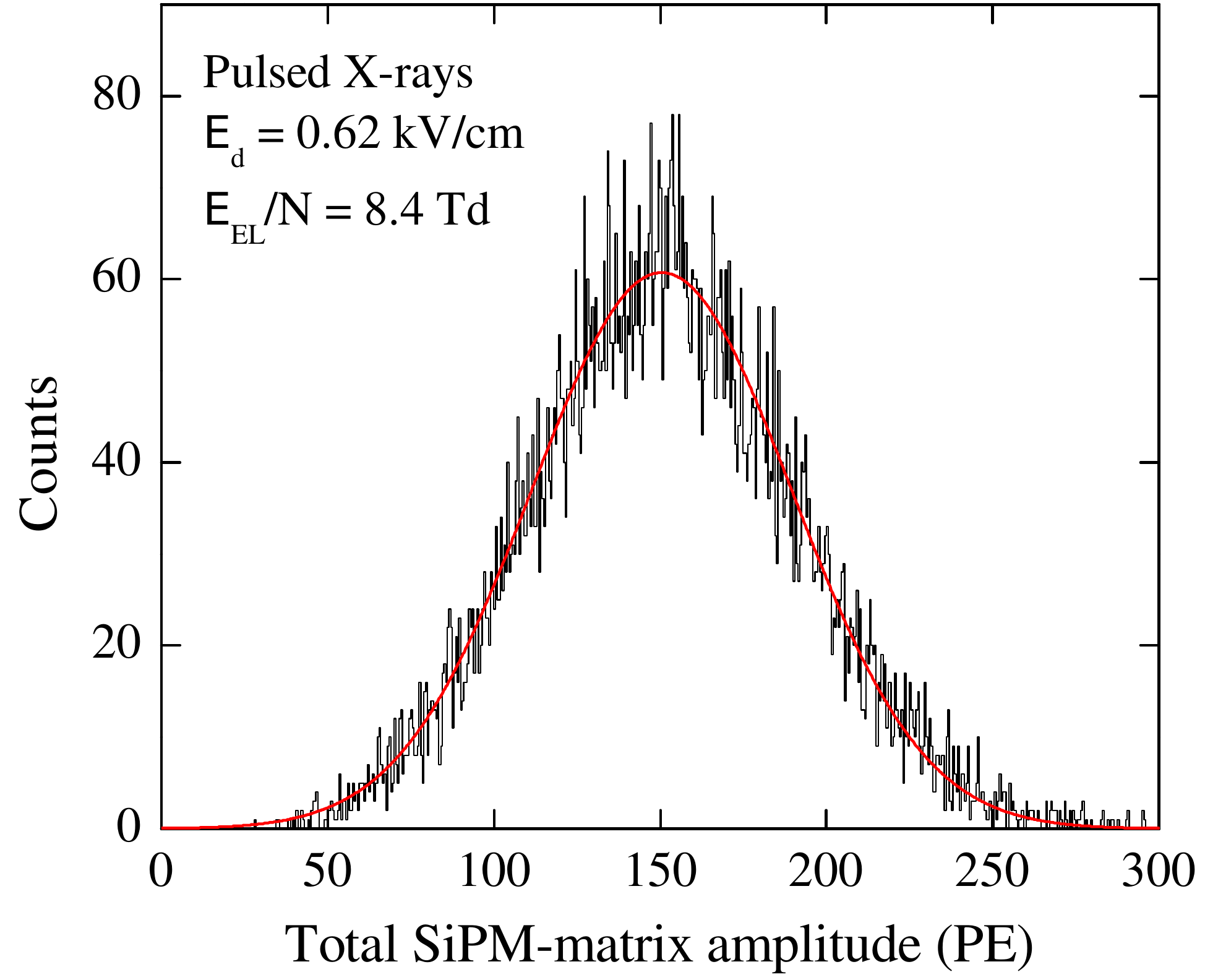}}
	\caption{
		Direct SiPM-matrix readout: amplitude spectrum of the total SiPM-matrix signal obtained with pulsed X-rays and 2 mm collimator.
		Red curve is fit by Gauss function.
		The data were obtained at the maximum reduced EL field, of 8.4~Td}
	\label{image:fig_171123_x_ray_20kV_2mm_SiPM_spectrum}
\end{figure}

Figure~\ref{image:fig_171123_x_ray_20kV_2mm_SiPM_spectrum} shows an example of the amplitude spectrum of the total SiPM-matrix signal, at the maximum EL field (compare to Figure~\ref{image:fig_171123_SiPM_48V_Cd_Npe_spec_82keVcuts_6mm_20kV}).

\begin{figure}[ht!]
	\center{\includegraphics[width=0.6\columnwidth]{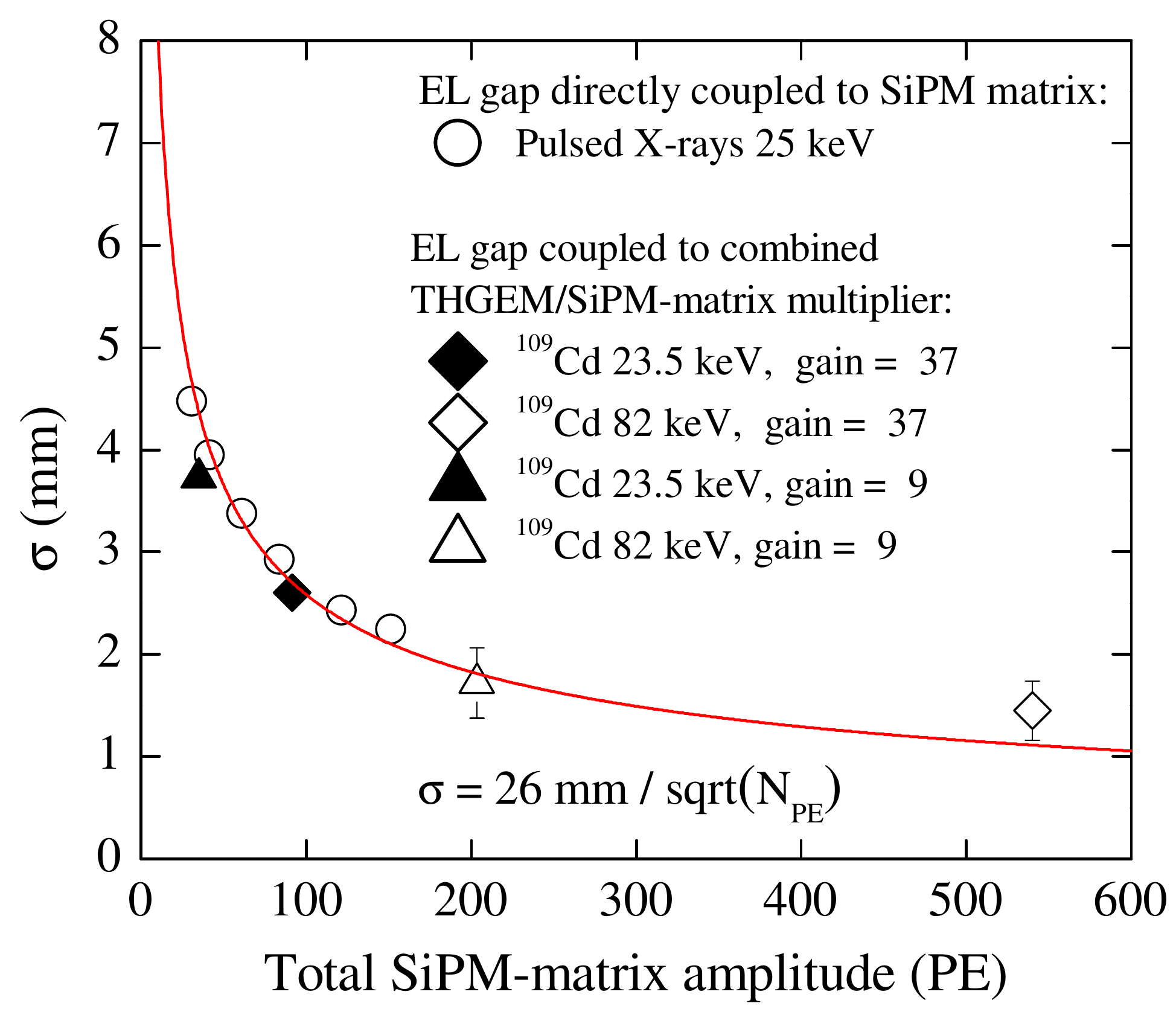}}
	\caption{Summary of position resolution results obtained in the two-phase detector for the direct SiPM-matrix and THGEM/SiPM-matrix readout. Shown is the position resolution (standard deviation) as a function of the total number of photoelectrons recorded by the SiPM matrix. Red curve is the fit by inverse root function using all data points}
	\label{image:fig_coord_res_vs_Npe_xray_Cd_THGEM_2mm_MC}
\end{figure}

Figure~\ref{image:fig_171123_coorg_2Dgr_xray_2mm_MC_20kV}, \ref{image:fig_171123_XYcoorg_xray_2mm_MC_20kV} and~\ref{image:fig_171123_x_ray_20kV_2mm_SiPM_spectrum} characterize the detector performance at the maximum EL field.
The position resolution and the average number of photoelectrons for lower fields were obtained in a similar way. These allow to define the dependence of the position resolution on the average total number of photoelectrons recorded by the SiPM matrix ($N_{PE}$): see Figure~\ref{image:fig_coord_res_vs_Npe_xray_Cd_THGEM_2mm_MC}.

The similar dependence, namely the position resolution as function of the total photoelectron number, was also obtained for the THGEM/SiPM-matrix readout.  
Here the $^{109}$Cd source was used instead of pulsed X-rays, to avoid the problems related to electronics saturation induced by high photon flux in the latter case. The procedure to measure the position resolution with the $^{109}$Cd source was generally similar to that with pulsed X-rays. The difference was that in the fit of $x_0(x_{exp})$ and $y_0(y_{exp})$ distributions the background due to Compton scattering of gamma-rays was taken into account (described by a wide Gauss function). The position resolution was measured for different $^{109}$Cd source energies, of 23.5~keV and 82~keV, and for different THGEM charge gains, of 9 and 37. The resulting dependence is shown 
Figure~\ref{image:fig_coord_res_vs_Npe_xray_Cd_THGEM_2mm_MC}.

Looking at the figure one may conclude that the position resolution does not depend on the readout concept: it has a universal character, depending only on the total photolectron number recorded by the SiPM matrix ($N_{PE}$), 
described by the inverse root function: 
\begin{equation}
\label{eq.Resolution} 
\sigma = 26\text{~mm} / \sqrt{N_{PE}} \, .
\end{equation}

This is surprising, since the readout geometry in both concepts is different. This universality might be due to the fact that in both readout concepts the THGEM1 electrode is used in front of the SiPM matrix, where THGEM1 holes act either as passive (light-transmitting) elements of an optical mask or as active (light-emitting) elements. Another possible explanation is that with a fairly large SiPM spacing in the SiPM matrix (1~cm), the difference in the distances to spatial regions where the light is produced for both readout concepts becomes insignificant.

\section{Discussion}\label{minimum_detection_threshold}

In this section we will try to estimate the detection thresholds in two-phase argon detectors with direct SiPM-matrix and THGEM/SiPM-matrix readout. The detection threshold is defined as the minimum energy, deposited by a scattered particle in the noble-gas liquid, that can be detected by the detector. There are two types of particle scattering: that of electron recoil, induced by gamma-ray and minimum ionizing particle scattering, or resulting from beta decays, such as by $^{39}$Ar, and that of nuclear recoil, induced by neutron and dark matter particle (WIMP) scattering. Their recoil energies are expressed in terms of keVee and keVnr respectively \cite{Chepel2013}.

The EL gap yields obtained in this work, of 0.022~PE/e$^-$ for direct SiPM-matrix readout and 0.65~PE/e$^-$ for THGEM/SiPM-matrix readout, can be significantly increased by optimizing the two-phase detector. Firstly, in direct SiPM-matrix readout, the THGEM1 anode with optical transparency (combined with the angle reduction factor) of only $0.28 \cdot 0.40 = 0.11$ can be replaced by the transparent electrode with ITO coating. Secondly, the sensitive area of the SiPM matrix can be increased from the current 36\% (see Figure~\ref{image:fig_SiPM_11x11_matrix_photo}) to about 90\%, pushing the SiPMs close to each other.
Consequently in optimized conditions, the amplitude yield can be increased up to about 0.5~PE/e$^-$ for direct SiPM-matrix readout (at EL reduced field of 8.4~Td) and up to 1.6~PE/e$^-$ for THGEM/SiPM-matrix readout (at THGEM1 charge gain of 37).

The detection threshold of an S2 signal depends on the pulse shape and dark count rate. For certainty, let the detection threshold (in terms of the photoelectron number) be 10~PE. Indeed, this value is large enough  in terms of the position resolution (of about 1~cm) and energy resolution (of about 30\% assuming Poisson statistics). Now we can calculate the minimum number of electrons drifting in the EL gap, corresponding to 10~PE signal at the SiPM matrix, using the amplitude yields of the previous paragraph. It amounts to 20~e$^-$ and 6.2~e$^-$ for direct SiPM-matrix and THGEM/SiPM-matrix readout respectively.

For a rough estimate of the energy thresholds, we used the ionization yield in liquid argon measured in~\cite{Joshi} at low  energies (around several keVs):  it is 10~e$^-$/keVee for electron recoils and 3.6~e$^-$/keVnr for nuclear recoils,  at a drift field of 0.24~kV/cm (which is close to that used in DarkSide-50 experiment~\cite{Agnes2015}).
Here we implied a full 100\% electron transmission through THGEM0 electrode, since that can be easily reached by just increasing the voltage applied across THGEM0~\cite{Bondar2019_field_sim}.
The appropriate detection thresholds are presented in Table~\ref{table:Detection threshold}.

\begin{table}
	\centering
	\caption{Detection thresholds, corresponding to 10~PE signal at the SiPM matrix, that can be achieved under optimal conditions for alternative readout concepts of the two-phase argon detectors with 1.8~cm thick EL gap, expressed in drifting electrons in the EL gap ($e^-$) and in deposited energy in liquid Ar for electron (keVee) and nuclear (keVnr) recoils,  at a drift field in liquid Ar of 0.24~kV/cm.  Also shown are the EL gap yields.}
	\label{table:Detection threshold}
	\begin{tabular}{@{\extracolsep{\fill}}|ccccc|@{}}
		\hline
		\multirow{2}{*}{\makecell{Readout concept \\ }} & \multicolumn{3}{c}{\makecell{Detection \\ threshold for 10 PE}} & {\makecell{EL gap \\ yield}} \\ 
		& (e$^-$) & (keVee) & (keVnr) & (PE/e$^-$)\\
		\hline
		\rule{0pt}{7ex}
		\makecell{Direct SiPM-matrix \\ readout \\ (1.8~cm thick EL gap,\\ $\mathcal{E}_{EL} / N$ = 8.4~Td)} & 20 & 2 & 5.6 & 0.5 \rule{0pt}{7ex}\\ 
		\makecell{THGEM/SiPM-matrix \\ readout \\ (THGEM gain = 37)} & 6.2 & 0.6 & 1.7 & 1.6 \rule{0pt}{7ex}\\
		\hline
\end{tabular}
\end{table}

These values should be considered as just indicative.
In particular the detection threshold for nuclear recoils for direct SiPM-matrix readout is of the order of 6~keVnr, which is enough to search WIMPs with masses above 10 GeV. For THGEM/SiPM-matrix readout the threshold is a factor of 3 lower, of the order of 2~keVnr, which is already close to that of DarkSide-50 experiment~\cite{Agnes2018}. 
Moreover, it can be further decreased, by increasing the THGEM charge gain, for example by using the double-THGEM multiplier~\cite{Buzulutskov2012,Buzulutskov2020}.

As an example, let us consider the detector response to a 1~MeV signal induced by solar neutrino interaction, in particular when recording the CNO neutrinos as proposed in~\cite{Aalseth2018}. Taking into account the ionization yield from~\cite{Shibamura1975} and the EL gap yield from Table~\ref{table:Detection threshold}, for 1~MeV signal the number of drifting electrons and photolectrons for the THGEM/SiPM-matrix readout would exceed 12000 and 20000 respectively, which would provide a sufficient energy resolution according to Poisson statistics, of about 1\% root-mean-square. 

Let us evaluate now the position resolution properties of the SiPM-matrix readout in comparison with a PMT-matrix readout. Table~\ref{table:spatial_resolution_for_different_detectors} compares the position resolution at a certain photoelectron number, reported in different dark matter search experiments using PMT-matrix readout, to that obtained in this work and extrapolated to the given photoelectron number using Eq.~\ref{eq.Resolution}.

\begin{table}
	\centering
	\caption{Position resolution of two-phase detectors in $x$,~$y$ plane extrapolated from that obtained in this work (using SiPM-matrix readout with 1 cm channel pitch) in comparison with that reported in dark-matter search experiments (using PMT-matrix readout).
		Also shown is the data for dual-phase Xe TPC with SiPM-matrix readout~\cite{Baudis2020}.}
	\label{table:spatial_resolution_for_different_detectors}
	\begin{tabular}{@{\extracolsep{\fill}}|ccc|@{}}
		\hline
		Experiment & \makecell{Reported position \\ resolution \\ } & \makecell{Position resolution \\ expected for \\ SiPM matrix with \\  1 cm channel pitch} \\
		\hline
		\rule{0pt}{5ex}
		\makecell{This work} & \makecell{ $\sigma = \frac{26\text{ mm}}{\sqrt{N_{PE}}}$ }& - \\  
		\makecell{LUX~\cite{Akerib2018_LUX}}& \makecell{ $\sigma = \frac{75\text{ mm}}{\sqrt{N_{PE}}}$ }& $\sigma = \frac{26\text{ mm}}{\sqrt{N_{PE}}}$ \rule{0pt}{5ex} \\
		\makecell{XENON100~\cite{Aprile2012_XENON100}} & \makecell{ $\sigma (46000\text{~PE})$ \\ = 3 mm } & 0.12~mm \rule{0pt}{5ex} \\
		\makecell{XENON1T~\cite{Aprile2017_XENON1T}} & \makecell{ $\sigma (200\text{~PE})$ \\ = 20 mm } & 1.8~mm \rule{0pt}{5ex} \\
		\makecell{DarkSide-50~\cite{Agnes2018_DS50_532day}} & \makecell{ $\sigma (20000\text{~PE})$ \\ = 6 mm } & 0.18~mm \rule{0pt}{5ex}  \\
		\makecell{Dual-phase Xe TPC \\ with SiPM matrix~\cite{Baudis2020}} &  \makecell{ $\sigma (1000\text{~PE})$ \\ = 1.5 mm } & 0.82~mm \rule{0pt}{5ex}  \\
		
		\hline
	\end{tabular}
\end{table}

This extrapolation can only be considered indicative. Nevertheless it allows to conclude that the position resolution of the SiPM-matrix readout is always superior to that of PMT-matrix readout, by a factor varying from 3 to more than an oder of magnitude. This superiority can be explained by a decrease in the channel pitch, from 3 inches in the case of PMT-matrix to 1~cm in the case of SiPM-matrix.

The obvious application of superior position resolution is the precise x-y fiducialization, which can help reducing background from the TPC wall with minimal loss of fiducial mass. 
In addition, at sufficiently high event energy, high readout granularity can enable resolving two nearby scattering vertices with the same drift time.

In this work we considered the issue of S2-only signal detection, thus overlooking the problem of S1 signal detection. In this case, the background rejection should be similar to that applied in dark matter search experiments operated in S2-only detection mode~\cite{Agnes2018,Aprile2019_XENON1T}, in particular using effective passive TPC shielding and proper MC simulation.

Finally, one should mention the possibility to record the S1 signal directly, without TPB, i.e. similarly to that of S2 in this work: using the effect of primary scintillations in the visible range in liquid Ar observed elsewhere~\cite{Buzulutskov2011,Alexander2016}, albeit at lower light yield compared to ordinary (VUV) scintillation.

\section{Conclusions}

In this work, we have for the first time demonstrated two alternative techniques of the SiPM-matrix readout of two-phase argon detectors, using electroluminescence (EL) in the visible and NIR range induced by either neutral bremsstrahlung (NBrS) or avalanche scintillation. 

In the first technique, the EL gap was directly read out by the SiPM matrix. In the second technique, the EL gap was read out via combined THGEM/SiPM-matrix multiplier, the THGEM being operated in electron avalanche mode. 

The amplitude yield was measured for these readout techniques: under optimal conditions it would amount to about 0.5~PE/e$^-$ and 1.6~PE/e$^-$ for the direct SiPM-matrix and THGEM/SiPM-matrix readout respectively. This allowed to assess the detection threshold in two-phase argon detectors for dark matter search: for nuclear recoils it was estimated to be of the order of 6~keVnr and 2~keVnr, respectively.

Using the SiPM matrix with 1~cm channel pitch, we obtained the highest position resolution ever measured for two-phase detectors with an EL gap: $\sigma = 26\text{~mm} / \sqrt{N_{PE}}$. 

Unlike the ``standard'' optical readout of two-phase TPCs (in the VUV), both alternative readout techniques allow to operate without TPB, which is particularly valuable for large-scale detectors. 
In particular, the results of this study were intended for use in the DarkSide-20k experiment: the alternative readout techniques might be considered as backup solutions, in case issues with TPB instability over time or non uniformity over large areas should become problematic.

There is another possible application of the NBrS EL signal in the DarkSide experiment. Due to its fast nature, its pulse width can be used to accurately measure the EL gap thickness even under current experimental conditions (i.e. using WLS), provided that the EL gap operates at lower fields (below 4~Td), where the S2 slow component of ordinary EL disappears and thus does not interfere with measurements.

Finally, it should be emphasized that, to the best of our knowledge, this is the first practical application of the NBrS effect in detection science.

\acknowledgments
The minor part of the work, regarding the study of SiPM matrices in sections~3, was supported by Russian Foundation for Basic Research (project no. 18-02-00117). The major part of the work, including  the results on EL yield and SiPM-matrix readout in sections 4-10, was supported by Russian Science Foundation (project no. 20-12-00008), The work was done within the R\&D program for the DarkSide-20k experiment.


\bibliographystyle{spphys_modified}
\bibliography{mybibliography}

\end{document}